\documentclass[twocolumn,tighten,astrosymb,twocolappendix]{aastex7}

\usepackage{amsmath,amssymb,amstext}
\usepackage{subfigure}
\usepackage{float}
\usepackage[version=4]{mhchem}
\usepackage{multirow}
\usepackage[utf8]{inputenc}
\usepackage{lmodern}
\usepackage[T1]{fontenc}
\usepackage{mathtools,nccmath}
\usepackage{environ}
\usepackage{natbib}
\usepackage{hyperref}
\usepackage{balance}

\usepackage{times}
\usepackage{epsfig}
\usepackage{graphicx}
\usepackage{epstopdf}
\usepackage{bm}
\usepackage{color}
\usepackage[english]{babel}
\usepackage{comment}

\usepackage{xcolor}
\usepackage{soul}

\newcommand{\s}{  s$^{-1}$}

\newcommand{\syn}{\mathrm{syn}}

\usepackage{tikz}
\usetikzlibrary{arrows.meta, positioning}
\usepackage{enumitem}

\begin{document}
	
	\title{\large Hadronic Processes, Plasma Evolution and Neutrino Emission in Magnetic Towers of Neutron-Star Merger Remnants}

	\author{Rostom Mbarek}
	\email[show]{rmbarek@princeton.edu}
	\affiliation{Department of Astrophysical Sciences, Princeton University, Princeton, NJ 08544, USA}
	
	\author{Jiaxi Wu}
	\affiliation{TAPIR, Mailcode 350-17, California Institute of Technology, Pasadena, CA 91125, USA}
	\email{jiaxiwu@caltech.edu}
	
	\author{Elias R. Most}
	\email{emost@caltech.edu}
	\affiliation{TAPIR, Mailcode 350-17, California Institute of Technology, Pasadena, CA 91125, USA}
	\affiliation{Walter Burke Institute for Theoretical Physics, California Institute of Technology, Pasadena, CA 91125, USA}
	
	\begin{abstract}

	Binary neutron star mergers can form short-lived magnetar-like remnants whose magnetically dominated polar towers reach $B\sim10^{15}$--$10^{16}\,\mathrm{G}$, but the microphysical composition of these outflows remains poorly understood. Combining tower geometries from GRMHD simulations with an analytic treatment of QED and hadronic processes, we argue that magnetic reconnection is the most viable particle acceleration channel in this strongly radiative regime, where the current sheets thin to collisionless scales. Purely leptonic pair loading---including resonant inverse Compton scattering of soft photons---is bottlenecked by rapid pitch-angle damping and the tendency of one-photon magnetic conversion to populate low Landau levels. Once protons reach mildly relativistic energies ($\gamma_p\gtrsim1.3$), however, inelastic proton-proton ($pp$) collisions inject large-pitch-angle pions that drive $\pi^0\to2\gamma\to e^\pm$ cascades with multiplicity $\mathcal{M}_{\rm cas}\simeq4$ at $B=10^{15}\,\mathrm{G}$, supplying the perpendicular momentum the leptonic channel cannot maintain. This hadronic route dominates pair loading and channels most of the dissipated magnetic energy into the $e^\pm$ population that could power the nonthermal emission emerging at larger radii. Charged-pion decay, modulated by $\pi^\pm$ synchrotron cooling, further seeds a nonthermal neutrino tail up to $\sim 300\,(\sigma_p/5)\,\mathrm{MeV}$, spectrally distinct from the thermal cooling burst and detectable from sources within $\sim 100\,\mathrm{kpc}$.

    \end{abstract}
	\section{Introduction}
	\label{sec:intro}
	
	Neutron star mergers can produce some of the most extreme magnetic fields in the universe \citep[e.g.,][]{kiuchi+15,giacomazzo+15}. For low-mass mergers where a long- or intermediate-lived neutron-star remnant is formed \citep[e.g.,][]{hotokezaka+13}, numerical relativity simulations coupled to general-relativistic magnetohydrodynamics (GRMHD) have painted the following picture: Small-scale dynamo amplification during merger \citep{Price:2006fi,Kiuchi:2015sga,Kiuchi:2017zzg,Aguilera-Miret:2021fre,Palenzuela:2021gdo,Chabanov:2022twz,Gutierrez:2026ngt}, and subsequent mean-field dynamo amplification in the differentially rotating \citep{Hanauske:2016gia} remnant \citep{Most:2023sft,combi+23,Kiuchi:2023obe,Most:2023sft} will lead to the production of magnetar-like magnetic fields. Recent numerical relativity simulations demonstrate that the field formed during merger can buoyantly break out from either the star \citep{Most:2023sft,combi+23,Kiuchi:2023obe} or the disk \citep{Musolino:2024sju} (see also Ref. \citet{Fields:2025afm}), and then assemble a large-scale magnetic tower structure attached to the star \citep{2003MNRAS.341.1360L}. This tower formation may help explain gamma-ray burst precursors, and may even contribute to the prompt emission, in a proposed subpopulation of short-duration bursts \citep{Gottlieb:2023sja}. One challenge in interpreting and predicting burst properties, and high-energy transients from these towers more generally, is their uncertain microphysical composition. This composition is not captured by current numerical models and remains poorly explored analytically, though see \citealt{harding+06} for related work in the context of magnetars. Magnetic towers are likely sites of strong magnetic dissipation, pair loading, and high-energy emission. 
	Collectively, current simulations support characteristic tower field strengths of $10^{15}$--$10^{16}\,\mathrm{G}$, with magnetization $\sigma = B^2/(4\pi n_p m_p c^2)\sim5$--$10$ and asymptotic Lorentz factors $\Gamma\sim5$--$10$ \citep{kiuchi+15,ruiz+16,ciolfi20,combi+23}, though these simulations have in part uncontrolled numerical floor uncertainties, see, e.g., \citet{Kalinani:2025itu}. Here $n_p$ ($m_p$) is the hadron number density (mass).
	
	While GRMHD simulations calculate the macroscopic background, they do not self-consistently determine the microphysical plasma composition. 
	Standard ideal GRMHD---even when augmented with approximate neutrino transport---does not include hadronic interactions or strong-field QED source terms that control pair creation. The lepton content sets the effective electron inertia and therefore the electron magnetization $\sigma_e \simeq (m_p/m_e)\sigma$ in a pure proton--electron plasma. 
	In this work, we take initial steps to alleviate this shortcoming by addressing the microphysical processes potentially active in magnetic towers formed in BNS. We do so using a hybrid approach: GRMHD-informed conditions for the macroscopic background and potential acceleration sites, supplemented by an analytic treatment of weak-interaction and QED processes to quantify pair loading.
	
	We assume the particle population at the tower base is thermal and ask whether protons can be boosted to mildly relativistic energies despite strong radiative losses. The most viable acceleration channel in this system is magnetic reconnection (Section~\ref{sec:acceleration}). Pair creation near the base could lower $\sigma_e$ by increasing lepton inertia, while neutrino annihilation $\nu\bar\nu\to e^+e^-$ enhances pair energy deposition at the order-unity level, with a factor of $\sim2$ increase for long-lived massive remnants \citep{perego+17}. Electron synchrotron is unlikely to provide efficient pair injection in the inner tower, since large pitch angles are rapidly damped and magnetic conversion tends to populate low Landau states \citep{harding+06,beloborodov+07,gill+14,thompson+14}.
	
	Motivated by these considerations, we argue that mildly relativistic protons play a central role through $pp\to\pi^0\to2\gamma$ followed by $\gamma B\to e^\pm$. This channel can surpass QED processes in pair injection because neutral-pion decay produces photons with a quasi-isotropic angular distribution in the center-of-mass frame, enabling pairs to populate high Landau levels and strengthening electromagnetic cascading.
	Charged-pion synchrotron provides an additional $\gamma$-ray source. For $B\gtrsim B_{\rm syn,\pi^\pm}$ defined by $t_{\rm syn,\pi^\pm}=t_{\rm dec,\pi^\pm}$ (Appendix~\ref{app:pion_decay_compete}), synchrotron damps the perpendicular momentum of $\pi^\pm$ before decay, so the cutoff in Eq.~\eqref{eq:Enu_max} of Appendix~\ref{app:neutrino_detect} becomes an upper envelope set by the residual parallel Lorentz factor $\gamma_{\parallel,\pi}$ rather than by $\gamma_\pi\simeq\sigma_p \simeq \sigma$. The pion (and subsequent muon) decays still contribute a nonthermal neutrino component at $\sim$100 MeV, spectrally distinct from the $T_\nu\sim 5$--$10\,\mathrm{MeV}$ thermal cooling burst of the remnant \citep{Kyutoku:2017wnb, Cusinato:2021zin}.
	
	The paper is organized as follows. We identify reconnection as the likely dominant proton acceleration mechanism (Section~\ref{sec:acceleration}), characterize the cooling environment (Section~\ref{sec:cooling}), analyze QED and hadronic pair-loading channels (Section~\ref{sec:pairs}), and estimate the injected power budget (Section~\ref{sec:power}). Supporting derivations are given in the Appendices.
	
	\section{Particle Acceleration in Post-Merger Magnetic Towers}
	\label{sec:acceleration}
	
	The ultra-strong magnetic fields in post-merger towers enforce extremely small
	pitch angles, $\alpha\ll 1$. Sustained acceleration in this limit is generally challenging because many standard mechanisms rely on appreciable transverse
	motion, repeated scattering, or large-amplitude fluctuations, all of which tend to increase $\alpha$ and quickly amplify radiative losses. Shock acceleration is
	sometimes invoked as a driver of particle energization in magnetized polar outflows and proto-jets \citep[e.g.,][]{piran+04}, but it is not viable
	under the conditions considered here. In jet-like systems, the magnetic field is predominantly toroidal and therefore largely perpendicular to the flow direction, so shocks propagating along the jet axis are expected to be perpendicular shocks \citep[e.g.,][]{sironi+13}. Particle-in-cell simulations of relativistic perpendicular shocks show that Fermi acceleration is efficient only for magnetizations $\sigma \lesssim 10^{-3}$ in pair plasmas
	and $\sigma \lesssim 3\times 10^{-5}$ in electron--ion plasmas
	\citep{sironi+13,sironi+15c}. 
	Viable channels must energize particles predominantly along the magnetic field. Two candidates are relevant:

    \textit{Curvature acceleration} is in principle suited to $\alpha\ll1$, since particles gain energy while moving along magnetic field lines \citep[e.g.,][]{vega+24}. Efficient energization, however, requires particles to traverse sufficiently curved, time-dependent magnetic structures. Its efficiency is therefore controlled by the curvature radius $R_c$. In BNS towers, this requirement becomes severe: the relevant field gradients would have to develop on scales as small as $\sim10^{-3}\,\mathrm{cm}$ (Appendix~\ref{app:curv_drift_scales}), which appears implausibly small. Physically, a relativistic, nearly field-aligned particle drifts across field lines by only $d_\perp\sim R_c/\gamma\beta_\parallel\sim10^{-3}\,\mathrm{cm}$ over a curvature time, so any accelerating structure sampled by this drift must vary on a comparably tiny transverse scale.

	\textit{Magnetic reconnection} provides a natural low-pitch-angle acceleration
	channel: in thin current sheets, the non-ideal electric field can energize
	particles directly along the reconnecting layer, without requiring large
	gyration angles. Kinetic studies show that relativistic reconnection remains
	an efficient accelerator even when radiative losses are dynamically important.
	Radiative PIC simulations find that cooling compresses plasma and magnetic
	flux into dense plasmoids \citep{schoeffler+19,schoeffler+23,hakobyan+23b}, while 3D
	strong-cooling calculations show that the current-sheet and plasmoid
	structure is highly sensitive to the cooling efficiency and can still
	accelerate ions efficiently \citep{chernoglazov+23}. Complementary
	radiative-MHD work shows that cooling can reshape current-sheet formation
	itself, modifying the sheet length and reconnection rate \citep{chowdhry+25}.
	For magnetar-strength fields, a fully self-consistent treatment may also
	require QED vacuum-polarization corrections to Maxwell's equations
	\citep{alawashra+25}.
	
	\begin{figure}[h!]
		\centering
		\includegraphics[width=\linewidth]{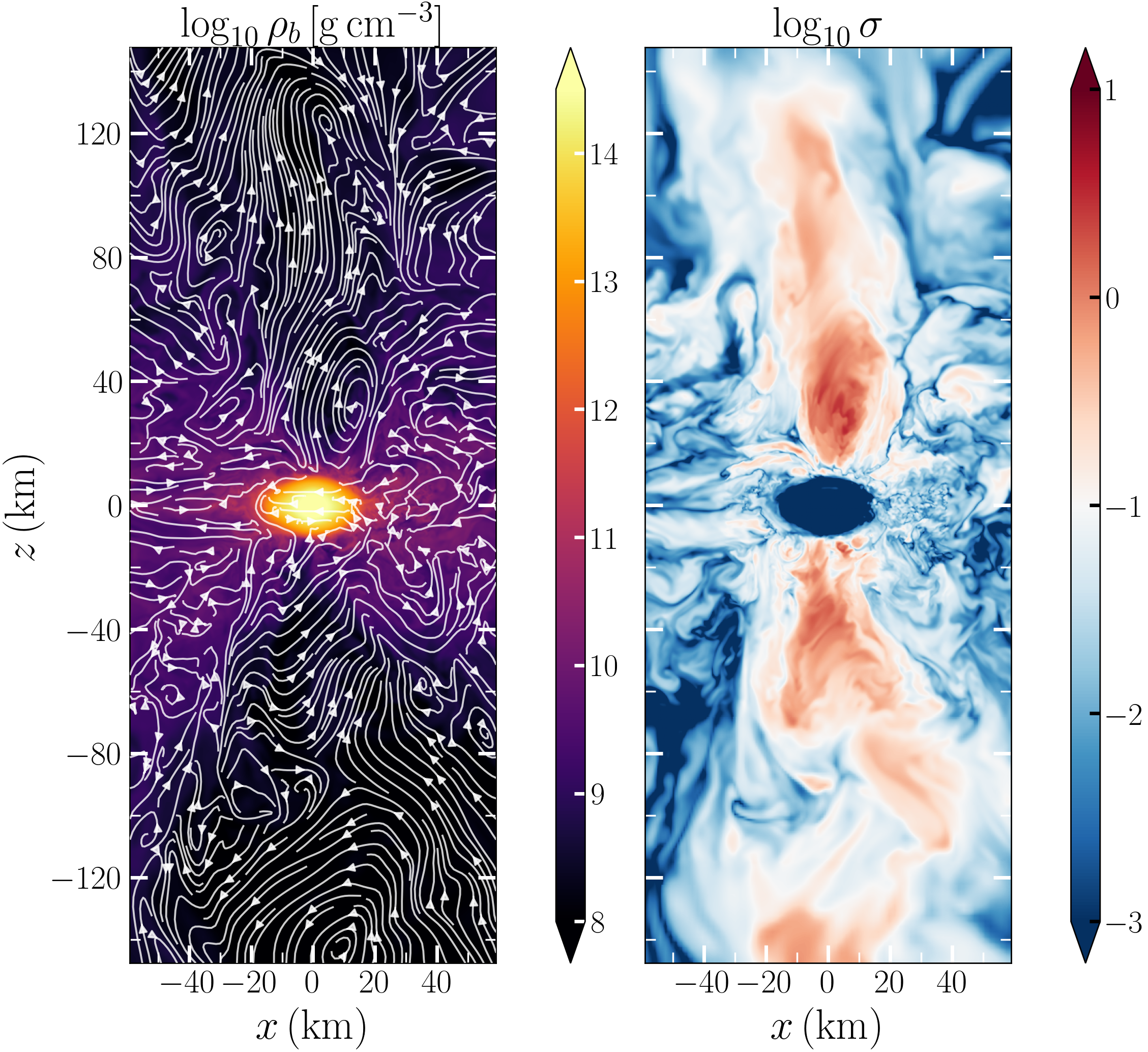}
		\caption{Meridional slice of the neutron-star merger remnant at $t\sim60\,\mathrm{ms}$ after merger. Left: baryon mass density $\rho_b$ (cgs) with magnetic field lines plotted on top. Right: magnetization $\sigma=b^2/(4\pi\rho_b c^2)$. A mildly relativistic magnetically dominated polar tower with $\sigma\gtrsim 1-10$ extends to $z\sim100\,\mathrm{km}$.}
		\label{fig:fig1_global_property}
	\end{figure}
	
	To locate candidate reconnection sites in a realistic magnetic tower geometry, we analyze a BNS merger simulation performed with the \texttt{Frankfurt/IllinoisGRMHD (FIL)} code \citep{Most:2019kfe,Etienne:2015cea} using initial data from the \texttt{LORENE} library \citep{Gourgoulhon:2000nn}. The two neutron stars are modelled using the DD2 equation of state \citep{Hempel:2009mc} with baryonic masses $1.27\,M_\odot$ and $1.43\,M_\odot$ (mass ratio $q=0.9$) at an initial separation of $35\,\mathrm{km}$. The mesh uses up to 9 refinement levels with a finest grid spacing of $214\,\mathrm{m}$. Because the numerical resolution is insufficient to fully capture the Kelvin--Helmholtz instability or MRI-driven dynamo directly, an effective mean-field dynamo is switched on post-merger to amplify the magnetic field \citep{Most:2023sme}. This model has been calibrated against high-resolution simulations in terms of Poynting flux and mass-density dependence (e.g., \citet{Kiuchi:2015sga,Kiuchi:2023obe}). The merger leaves a long-lived hypermassive neutron star (HMNS) remnant that launches a magnetically dominated polar tower. 
    The overall properties of the tower are comparable between different numerical simulations \citep{Most:2023sft,Kiuchi:2023obe,Combi:2023yav,curtis+24}, though the details of how the near-star geometry may be mildly affected by the precise launching mechanism \citep{Musolino:2024sju}. 
	
\begin{figure}
		\centering
		\includegraphics[width=\linewidth]{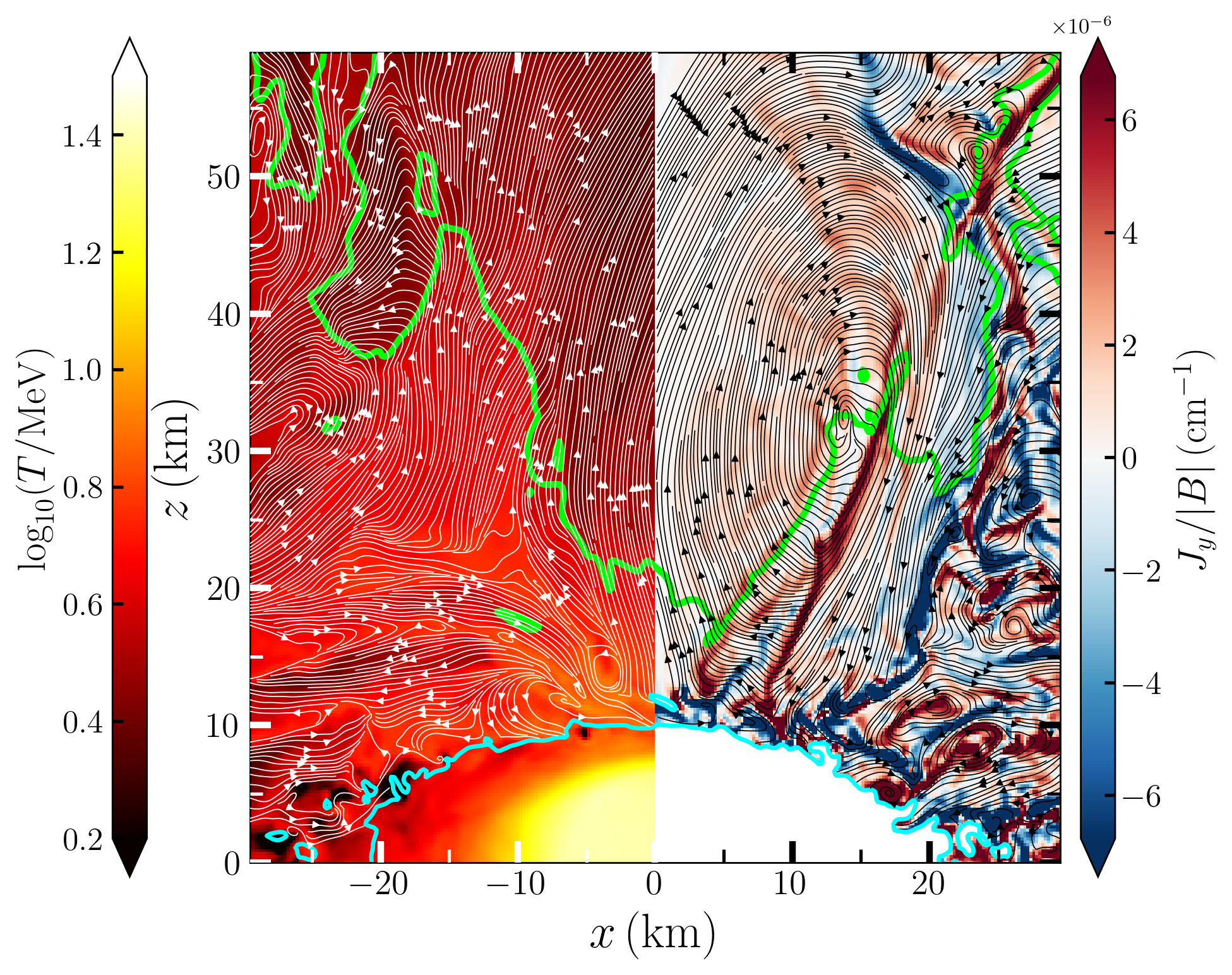}
		\caption{{\it (Left)} Map of temperature, $T$, and {\it (Right)} the out-of-plane current density normalized to the magnetic field strength $\boldsymbol{J}_y/|\boldsymbol{B}|=(\nabla\times\boldsymbol{B})_y/|\boldsymbol{B}|$ in the basal region of the tower, with overplotted magnetic field lines. The cyan contour marks the HMNS boundary ($\rho\gtrsim 10^{11}\mathrm{g\,cm^{-3}}$), and the lime contour marks the magnetic tower region ($\rho\lesssim 10^{9}\mathrm{g\,cm^{-3}}$). Narrow current sheets indicate candidate reconnection sites for particle acceleration.}
		\label{fig:fig2_reconnection_site}
	\end{figure}

	Figure~\ref{fig:fig1_global_property} shows the HMNS at $t\sim60\,\mathrm{ms}$. Figure~\ref{fig:fig2_reconnection_site} shows both the temperature and the out-of-plane current density $J_y \propto(\nabla\times\boldsymbol{B})_y$ with overplotted field lines in the upper tower. We identify several candidate reconnection sites near the tower base through thin high-$|J_y|$ layers coincident with sharp field-line reversals. These are identified morphologically rather than through direct measurement of a non-ideal electric field; nevertheless, they suggest that the tower provides viable sites for the mildly relativistic proton acceleration considered here. Based on particle-in-cell (PIC) results, we assume that particles can be accelerated up to energies of order $\sigma m_p c^2$ per particle \citep[e.g.,][]{sironi+14a}, consistent with the fraction of available magnetic power accessible at reconnection sites.
	
	The GRMHD simulation does not capture the kinetic acceleration process itself. It is used only to identify magnetized current-sheet-like regions in the tower. Once such sheets form, the particle energization is controlled mainly by the local magnetization\footnote{Neutrino-driven winds could in principle baryon-load the polar region and modify the magnetization. However, the magnetically dominated tower identified in Figure~\ref{fig:fig1_global_property} corresponds to the funnel where the wind is suppressed \citep{Musolino:2024sju}. Simulations including neutrino transport find that the wind is concentrated at latitudes off the rotation axis, while the polar column is magnetically dominated with $\sigma \geq 1$ \citep[e.g.,][]{perego+17,combi+23}} $\sigma_p$ and field topology, both of which are well within the regime where PIC simulations of relativistic reconnection demonstrate efficient nonthermal acceleration (Appendix~\ref{app:coulomb}). In particular, kinetic studies with $\sigma\geq 1$ consistently find that a substantial fraction of the dissipated magnetic energy is deposited into particles energized up to $\gamma\sim\sigma$, with further acceleration producing a nonthermal power-law tail \citep[e.g.,][]{sironi+14a,liu+15,zenitani+07,kagan+15,cerruti+12,cerruti+14a,rowan+17,ball+18,mbarek+22}.

	\section{Cooling in the Post-Merger Tower}
	\label{sec:cooling}
	
	The BNS merger tower has $B\sim10^{15}$--$10^{16}\,\mathrm{G}$, well above the electron quantum critical field $B_Q=4.4\times10^{13}\,\mathrm{G}$, placing the leptonic component deep in the strong-field QED regime. By contrast, protons and pions remain effectively classical with respect to magnetic quantization: $B_{Q,p}=(m_p/m_e)^2B_Q\sim10^{20}\,\mathrm{G}$ and $B_{Q,\pi}=(m_\pi/m_e)^2B_Q\sim3\times10^{18}\,\mathrm{G}$, both far above the field strengths reachable in the core of the remnant neutron star.
	
	Figure~\ref{fig:timescales} summarizes the characteristic timescales. The inelastic $pp$ timescale is extremely short across the relevant field range (Appendix~\ref{app:pp_inelastic}), making $pp$ collisions a robust pion source for modest proton acceleration. Neutral-pion decay, $\pi^0\to2\gamma$, injects photons that are rapidly converted into pairs via $\gamma B\to e^\pm$ in the supercritical field. Charged pions can themselves radiate before decaying: at high $B$, the pion synchrotron time can become shorter than the pion decay time (Appendix~\ref{app:pion_decay_compete}), so that $\pi^\pm$ synchrotron becomes a major sink of charged-pion energy and an efficient secondary photon source. For pions injected with characteristic lab-frame pitch angles $\sin\alpha_\pi\simeq0.5$ at $\gamma_p\sim2$--$4$ (Appendix~\ref{app:pion-pitch}), this regime is reached for $B\gtrsim{\rm few}\times10^{15}\,\mathrm{G}$.

	Curvature cooling is not expected to control the energetics in the mildly relativistic regime. For curvature radii $R_c\sim10^5$--$10^6\,\mathrm{cm}$, the curvature-loss timescale $t_{\rm curv}\simeq3mc^3R_c^2/(2e^2\gamma^3)$ remains far longer than the dominant interaction and radiative timescales for $\gamma\sim\mathrm{few}$, confirming that curvature radiation is subdominant compared with $pp$ collisions and synchrotron/cyclotron cooling in the regime emphasized here.
	
	We further note that proton--proton Coulomb relaxation is subdominant by a factor $10$--$30$ compared to inelastic $pp$ in hot reconnection sheets ($T\sim 5$--$10\,\mathrm{MeV}$). The ratio is $B$- and $\sigma_p$-independent (Appendix~\ref{app:coulomb}). In short, $t_C/t_{pp}\simeq 1$ in the cool $T\sim 1\mathrm{MeV}$ bulk, so a Maxwellian background is maintained, while $t_C/t_{pp}\sim 10$–$30$ in the hot $T\sim 5$–$10\,\mathrm{MeV}$ reconnection sheets, so nonthermal protons deposit their energy into pions before being Coulomb-isotropized. Although $\lambda_C \ll L$ globally, we have $\lambda_C/d_e \gtrsim 10^4$ throughout the tower, implying that Coulomb scattering cannot regulate the dissipation once current sheets thin to kinetic scales (Appendix~\ref{app:coulomb}). In the plasmoid-mediated regime, macroscopic MHD sheets fragment into a hierarchy of progressively thinner sheets that reach the skin depth, at which point the dynamics transitions to collisionless reconnection \citep[e.g.,][]{uzdensky+10}.
	
	\section{Pair Loading in the Tower}
	\label{sec:pairs}
	
	Pair loading is essential because, in a compact highly magnetized outflow, much of the dissipated high-energy radiation is expected to be reprocessed into $e^\pm$ pairs rather than escaping directly, intrinsically linking the radiative output and plasma composition \citep{metzger+14}.
	
	\subsection{QED pair loading}
	\label{sec:qed_pairs}
	
	A natural question is whether pairs can be generated efficiently by QED processes alone, via cyclotron or synchrotron emission followed by $\gamma B\to e^\pm$. As shown in Appendix~\ref{app:cyclotron_main}, this channel is generally inefficient unless electrons maintain substantial pitch angles at emission, consistent with full cascade calculations in strongly magnetized neutron-star magnetospheres \citep{medin+10}.
	
	Two bottlenecks suppress this route in BNS towers. First, the threshold for one-photon magnetic conversion is $\epsilon\ge2/\sin\theta$ (ground-state channel), so photons emitted nearly along $\mathbf{B}$ have strongly suppressed conversion. In radiatively damped geometries where pitch angles are small, conversion is delayed or negligible. Second, even when fundamental cyclotron photons do convert, they can only do so via the ground-state $(n,n')=(0,0)$ channel, since the fundamental cyclotron energy $\epsilon_{\rm cyc}=\sqrt{1+2B'}-1$ is always 2 units below the $(0,1)$ threshold $1+\sqrt{1+2B'}$ (Appendix~\ref{app:cyclotron_main}). Ground-state pairs carry no transverse excitation and cannot amplify the cascade through synchrotron de-excitation, consistent with detailed cascade simulations at high field \citep{medin+10}. The magnetic-moment constraint imposes only an extremely small pitch-angle floor, $\sin\alpha_{\rm min}\sim\gamma\rho_0/R_c\lesssim10^{-17}\gamma$ for $B_0=10^{15}\,\mathrm{G}$ and $R_c\sim10^5\,\mathrm{cm}$ (Appendix~\ref{app:pitch-ang}). Thus, without continuous injection of perpendicular momentum, electrons can collapse to tiny pitch angles and become inefficient at driving $\gamma B$ conversion.
	
	Taken together, electron synchrotron/cyclotron emission alone is not a strong pair-loading pathway. Maintaining nonnegligible pitch angles---which would revive strong cyclotron de-excitation and $\gamma B$ conversion at high Landau levels---generally requires continuous perpendicular momentum injection, as can be provided by the hadronic $pp$ channel considered below.
	
	\begin{figure}[t!]
		\centering
		\includegraphics[width=\linewidth]{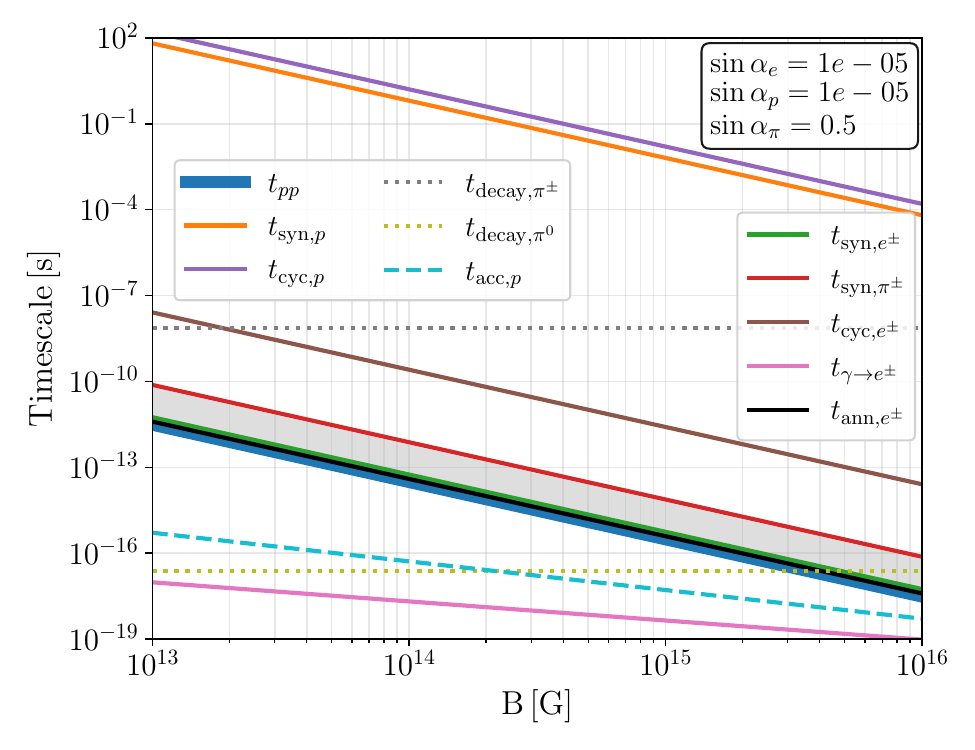}
		\caption{Characteristic timescales versus magnetic field strength $B$.
			Over the relevant field range, the inelastic $pp$ timescale, $t_{pp}$,
			lies below proton synchrotron, $t_{{\rm syn}, p}$, so hadronic collisions
			dominate over direct proton radiative losses for small proton pitch
			angles, $\alpha_p$. At high $B$, charged-pion synchrotron,
			$t_{\rm syn, \pi^\pm}$, becomes faster than pion decay,
			$t_{{\rm decay}, \pi}$. Once $B\gtrsim B_Q$, single-photon magnetic
			conversion is extremely rapid. Pair annihilation, $t_{{\rm ann}, e^\pm}$,
			remains slower than pair synchrotron cooling, $t_{{\rm syn}, e^\pm}$,
			so it does not quench the cascade. The shaded band shows the Coulomb
			collision time $t_C$ for $T=1$--$10\,\mathrm{MeV}$, spanning the bulk
			and current-sheet temperatures (Appendix~\ref{app:coulomb}); $t_C$
			tracks $t_{pp}$ closely in the cool bulk and lies an order of magnitude
			above it in the hot sheets. Inverse Compton is omitted because its pair-cascade efficiency is suppressed by the same Landau-level bottleneck that limits cyclotron pair injection (\S~\ref{sec:ic_angle_pair}), independent of the supplied power.}
		\label{fig:timescales}
	\end{figure}

	\subsection{Inverse-Compton pair loading}
		\label{sec:ic_angle_pair}
		
		Inverse Compton (IC) scattering can also upscatter photons into the pair-producing band, but in the inner tower IC pair injection is bottlenecked by the same Landau-level structure that limits the cyclotron channel of \S~\ref{sec:qed_pairs}. This bottleneck is closely related to the Landau-state dependence of magnetic pair creation and cascade development in magnetar-strength fields
		\citep{hu+19,hu+22,harding+25}. Electrons rapidly lose perpendicular momentum in the supercritical field and therefore move nearly along the local magnetic field. The seed photon energy relevant for magnetic Compton scattering is then the electron-frame energy $\epsilon_s'=\gamma_e\epsilon_s(1-\beta_e\cos\theta_s)$, where $\epsilon_s=E_s/(m_ec^2)$ and $\theta_s$ is the angle between the seed photon and the field-aligned electron. Counter-streaming photons are boosted, $\epsilon_s'\simeq2\gamma_e\epsilon_s$, whereas
		co-propagating photons are de-boosted, $\epsilon_s'\simeq\epsilon_s/(2\gamma_e)$, and cannot reach the cyclotron resonance for relativistic $\gamma_e$ unless $\epsilon_s$ already exceeds the Landau spacing. Resonant magnetic Compton scattering occurs at $\epsilon_s'\simeq\epsilon_{\rm cyc}\equiv\sqrt{1+2B'}-1$
		(Appendix~\ref{app:cyclotron_main}), giving
		\begin{equation}
			\gamma_{e,\rm res}
			\simeq
			\frac{\epsilon_{\rm cyc}}{\epsilon_s(1-\beta_e\cos\theta_s)} .
			\label{eq:gamma_res_ic_angle_main}
		\end{equation}
		For $B=10^{15}\,\mathrm{G}$ ($\epsilon_{\rm cyc}\simeq5.81$) and
		$E_s=100\,\mathrm{keV}$, this gives $\gamma_{e,\rm res}\simeq15$ for
		head-on photons and $\simeq30$ for photons incident at $90^\circ$, so mildly relativistic electrons can resonate with $\sim 100\,$keV seed photons in the tower.
		
		Pair creation by the upscattered photon is controlled by its transverse energy. The lowest-channel threshold,
		$\epsilon_{\rm IC}\sin\theta_{\gamma B}\gtrsim2$, populates ground-state $(0,0)$ pairs that carry no transverse excitation, while populating an excited Landau channel $(0,1)$---and thereby driving a subsequent de-excitation cascade---requires the larger transverse energy
		$\epsilon_{\rm IC}\sin\theta_{\gamma B}\gtrsim 1+\sqrt{1+2B'}$ (Eq.~\ref{eq:epsperp_thr01}). For resonant magnetic Compton scattering,
		the rest-frame emission is concentrated near $\theta'\sim\pi/2$ with
		photon energy $\epsilon'\simeq\epsilon_{\rm cyc}$. The lab-frame
		quantities are then $\epsilon_{\rm IC}\simeq\gamma_e\epsilon_{\rm cyc}$
		and $\sin\theta_{\gamma B}\simeq\gamma_e^{-1}$, so the transverse energy
		is independent of $\gamma_e$,
		\begin{equation}
			\epsilon_{\rm IC}\sin\theta_{\gamma B}\simeq\epsilon_{\rm cyc} ,
			\label{eq:eIC_sintheta_eps_cyc}
		\end{equation}
		exceeding the $(0,0)$ threshold of $2$ at $B\gtrsim B_Q$ but falling exactly $2$ units short of the $(0,1)$ threshold. Resonant IC photons therefore convert magnetically, but most likely via the ground-state branch, similar to the cyclotron-only bottleneck in Sec.~\ref{sec:qed_pairs}. 
		
		The same point applies to non-resonant IC. Populating the excited $(0,1)$ pair channel requires
		$\epsilon_{\rm IC}\sin\theta_{\gamma B}\sim\gamma_e\epsilon_s
		\gtrsim1+\sqrt{1+2B'}$, which implies $\epsilon_s'=\gamma_e\epsilon_s(1-\beta_e\mu_s)\gtrsim1$ for the relevant angles. Recoil and Landau-state structure are therefore important, so the interaction must be treated with the full
		magnetic Compton/QED kernel. The resulting photons are then processed by photon splitting and magnetic pair creation, as in the RICS cascade
		calculations of \citet{harding+25}.

	\subsection{Hadronic pair loading via proton--proton collisions}
	\label{sec:pp_pi0_pairs}
	
	If a persistent nonthermal proton tail reaches $\gamma_p\gtrsim\gamma_p^{\rm th}\simeq1.3$, the kinematic threshold for single $\pi^0$ production is met (Appendix~\ref{app:pp_inelastic}; \citealt{kafexhiu+14}). The $\pi^0$ mesons, produced with large pitch angles in the lab frame for $\gamma_p\lesssim5$ (Appendix~\ref{app:pion-pitch}; \citealt{kafexhiu+14,kamae+06}), promptly decay into two photons. The relevant kinematic point is that pions emitted isotropically in the $pp$ CM frame retain large lab-frame pitch angles ($\sin\alpha_\pi\sim 0.5$) only for $\gamma_p\lesssim 5$. At higher $\gamma_p$ the pion cone collapses by aberration, suppressing $\gamma B\to e^\pm$.
	
	The inelastic $pp$ rate is controlled by $t_{pp}^{-1}=n_p\sigma_{pp}\kappa_{pp}c$ with $\sigma_{pp}\sim(3$--$4)\times10^{-26}\,\mathrm{cm^2}$ and $\kappa_{pp}\sim0.5$ (Appendix~\ref{app:pp_inelastic}). A useful energy mapping is \citep{kelner+06,kamae+06}
	\begin{equation}
		E_\gamma^{(\pi^0)}\simeq b E_p,\qquad b\simeq 0.085,
		\label{eq:pi0_mapping_main}
	\end{equation}
	giving a total $\gamma$-ray fraction per inelastic collision $k_{\pi^0\gamma}=2b\simeq0.17$. Note that while \citet{kelner+06} and \citet{kamae+06} are appropriate at high proton energies, the near-threshold regime ($\gamma_p\lesssim5$) is best handled by the parametrization of \citet{kafexhiu+14}.
	
	\subsubsection{Cascade multiplicity and pair yield}
	\label{app:mcas_ypp}
	
	The effective cascade multiplicity $\mathcal{M}_{\rm cas}$ per injected primary photon is controlled by whether the initial $\gamma B\to e^\pm$ conversion injects pairs into excited Landau levels and how many subsequent de-excitation photons exceed the pair-creation threshold.
	
	For a photon with $\epsilon_\perp\equiv\epsilon\sin\theta\ge\epsilon_{\perp,\rm thr}^{(0,1)}\equiv1+\sqrt{1+2B'}$ (Eq.~\ref{eq:epsperp_thr01}), the primary conversion injects at least one lepton with $n>0$. For $\pi^0$-decay photons with $E_\gamma\sim10^2$--$10^3\,m_ec^2$ and approximately isotropic angular distributions ($\gamma_p\lesssim5$), this threshold is easily satisfied at $B=10^{15}$--$10^{16}\,\mathrm{G}$.
	
	A lepton injected into Landau level $n$ de-excites by emitting photons with spacing
	\begin{equation}
		\Delta\epsilon_n = \sqrt{1+2nB'}-\sqrt{1+2(n-1)B'}.
		\label{eq:landau_spacing}
	\end{equation}
	Since de-excitation photons are emitted predominantly at large angle to $\mathbf{B}$, a secondary pair is created whenever $\Delta\epsilon_n\gtrsim2$. The cascade multiplicity is
	\begin{equation}
		\mathcal{M}_{\rm cas}(B)\equiv 1+N_{\rm pair}^{\rm deex}(B),
		\label{eq:Mcas_def}
	\end{equation}
	where $N_{\rm pair}^{\rm deex}$ counts Landau de-excitation steps satisfying $\Delta\epsilon_n\ge2$. Evaluating Eq.~\eqref{eq:landau_spacing} for $B=10^{15}\,\mathrm{G}$ ($B'\simeq22.7$): $\Delta\epsilon_1\simeq5.81$, $\Delta\epsilon_2\simeq2.77$, $\Delta\epsilon_3\simeq2.13$, and $\Delta\epsilon_4\simeq1.80<2$. This yields $N_{\rm pair}^{\rm deex}=3$ and $\mathcal{M}_{\rm cas}\simeq4$.
	
	The cascade terminates rapidly because once a secondary photon converts, it preferentially populates the lowest Landau channels, producing ground-state pairs without a further energetic de-excitation branch.
	
	The mean number of $\pi^0$-decay photons per inelastic collision is $N_\gamma^{(\pi^0)}=2\langle n_{\pi^0}\rangle$, where \citep{kafexhiu+14}
	\begin{equation}
		\langle n_{\pi^0}\rangle = -6\times10^{-3}+0.237Q_p-0.023Q_p^2;
		\quad Q_p\equiv\frac{T_p-T_p^{\rm th}}{m_p}
		\label{eq:npi0_mean}
	\end{equation}
	where $T_p$ (energy) and $T_p^{\rm th}$ (threshold energy) are in the lab frame. If these photons convert before escape, the pair yield per inelastic collision is:
	\begin{equation}
		Y_{e^\pm/pp}(B)\simeq 2\langle n_{\pi^0}\rangle\,\mathcal{M}_{\rm cas}(B).
		\label{eq:Ypp_def}
	\end{equation}
	A proton confined long enough to undergo multiple inelastic collisions accumulates a total yield
	\begin{equation}
		N_{e^\pm/p}(\gamma_p,B)\simeq\sum_{i=0}^{N_{\rm int}-1}Y_{e^\pm/pp}(B,T_{p,i}),
		\label{eq:Yp_tot}
	\end{equation}
	where $T_{p,0}=(\gamma_p-1)m_pc^2$, $T_{p,i+1}\simeq(1-\kappa_{pp})T_{p,i}$, and the sum terminates when $T_{p,i}<T_p^{\rm th}$.
	
	\subsubsection{Synthesis}
	
\begin{figure}[t!]
		\centering
		\includegraphics[width=\linewidth]{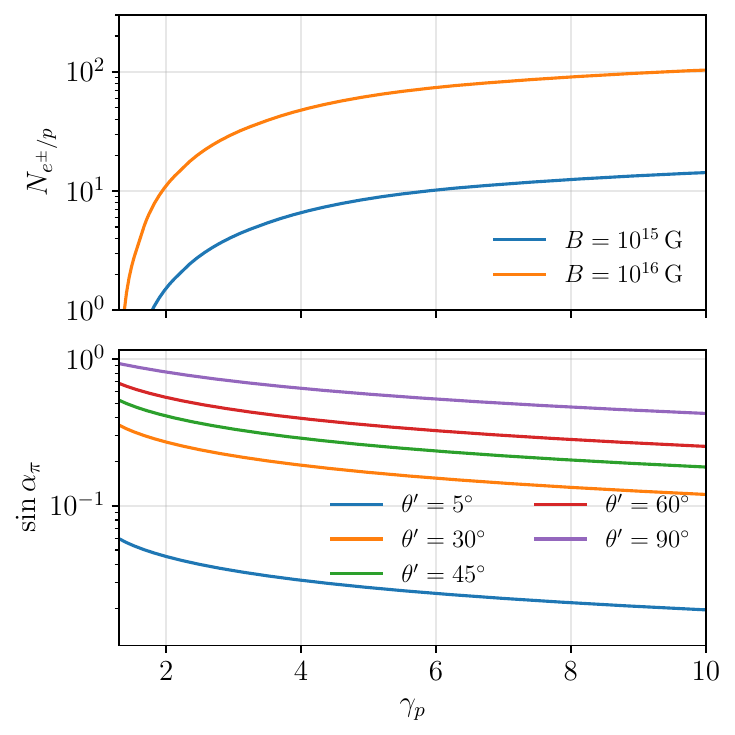}
		\caption{\textit{Upper panel}: Calorimetric pair yield per injected proton $N_{e^\pm/p}$ for different proton $\gamma_p$ for $B=10^{15}$ and $10^{16}\,\mathrm{G}$. These are upper limits; see text. \textit{Lower panel}: Lab-frame pion pitch angle $\sin\alpha_\pi$ for several pion CM emission angles $\theta'$ (Appendix~\ref{app:pion-pitch}). Increasing beaming at high $\gamma_p$ suppresses the effective pair yield relative to the calorimetric estimate.}
		\label{fig:yield}
	\end{figure}

	Figure~\ref{fig:yield} illustrates the key tradeoff. The upper panel shows the calorimetric pair yield per injected proton from Eq.~\eqref{eq:Yp_tot} for $B=10^{15}$ and $10^{16}\,\mathrm{G}$. The lower panel shows the lab-frame pion pitch angle $\sin\alpha_\pi$ for several CM emission angles (Appendix~\ref{app:pion-pitch}). As $\gamma_p$ increases, the pair yield grows because more energetic collisions produce more pions. However, the pion distribution becomes progressively more forward-beamed, reducing the magnetic conversion efficiency. Since the efficiency of $\gamma B\to e^\pm$ depends on the photon--field angle through $\chi_\gamma\propto(E_\gamma/m_ec^2)(B/B_Q)\sin\theta_B$ (Appendix~\ref{app:gamma_to_pairs_B}), the yields in the upper panel should be treated as calorimetric upper limits. The true yield is expected to be suppressed at large $\gamma_p$ by the beaming effect quantified in the lower panel. 
	
	\section{Injected Power from Proton--Proton Interactions}
	\label{sec:power}
	
	Understanding the energy partition in BNS postmerger towers is important
	because these polar outflows could contribute to short
	$\gamma$-ray burst diversity \citep{Gottlieb:2023sja}, and how dissipation channels energy into photons,
	pairs, or trapped secondaries directly affects the prompt radiative
	efficiency, composition, and angular structure of the emerging jet
	\citep{ciolfi20,Mosta:2020hlh,pavan+21,Kalinani:2025itu}.

	\paragraph{Calorimetric $pp$ regime}
	A central feature of the inner tower is that the inelastic $pp$ mean
	free path $\lambda_{pp} = (n_p \sigma_{pp} \kappa_{pp})^{-1}$ is
	tiny at all relevant field strengths. For $B = 10^{15}$\,G and $\sigma_p = 5$ one finds $\lambda_{pp} \sim 10^{-5}$\,cm and $t_{pp} \sim 10^{-16}$\,s, compared with an Alfv\'en crossing time $t_A = H/v_A \sim 3.6\times 10^{-4}$\,s over a tower height $H \simeq 100$\,km over which magnetar-level fields are present. The ratio $t_{pp}/t_A \lesssim 10^{-12}$ is roughly similar across the plotted field range. This means the $pp$ channel operates in the calorimetric limit: every nonthermal proton deposits its full kinetic energy into pions essentially at the point of creation. The sustainable hadronic power density is therefore supply-limited by reconnection dissipation, not by the instantaneous collision rate:
	\begin{equation}
		P_{\rm had}(B) \;=\; f_{\rm acc}\,\frac{U_B}{t_A},
		\label{eq:Phad_supply}
	\end{equation}
	where $U_B = B^2/(8\pi)$ and $f_{\rm acc}\simeq 0.1$ is a system-averaged efficiency. Equivalently, $f_{\rm acc}\sim (v_{\rm rec}/v_A)\,f_V\,(H/L_{\rm sheet})$, combining the local reconnection rate $v_{\rm rec}/v_A\sim 0.1$, an estimated volume-filling fraction of active current sheets $f_V\sim 0.1$ (Figure~\ref{fig:fig2_reconnection_site}), and the ratio of tower height to sheet length $H/L_{\rm sheet}\sim 10$. This estimate does not assume that every particle encounters a reconnecting sheet. Reconnection is instead taken to be the volumetric channel through which the tower dissipates its magnetic reservoir, while $f_{\rm acc}$ specifies the fraction of that dissipated power transferred to hadrons. For $B=10^{15}\,\mathrm{G}$ and a cylindrical tower of height $\simeq 100\,\mathrm{km}$ and radius $\simeq 10\,\mathrm{km}$, one finds $f_{\rm acc}U_BV \simeq 10^{47}\,\mathrm{erg}$ per Alfv\'en time. 
	

	Other consequences of this calorimetric limit are worth stating. First, nonthermal protons cannot free-stream out of the tower before undergoing inelastic $pp$ interactions: any proton-channel contribution to high-energy cosmic rays from this region is suppressed by $t_{pp}/t_{\rm esc}\lesssim 10^{-12}$. Second, the hadronic emission is spatially localized to regions in which reconnection is actively accelerating particles. 
	
	We also note that this opacity argument is not restricted to the innermost tower. It has recently been proposed \citep{farrar25} that ultra-high-energy cosmic rays (UHECRs) are accelerated in the turbulent magnetized outflow surrounding BNS merger remnants. Even taking the numerical atmosphere floor $\rho\sim 10^{3}\,\mathrm{g\,cm^{-3}}$ adopted in
	GRMHD simulations \citep[e.g.,][]{cook+25} as representative of the sheath surrounding the tower
	\citep{combi+23}, one finds $n_p\sim 6\times 10^{26}\,\mathrm{cm^{-3}}$ and an inelastic
	mean free path $\lambda_{pp}^{\rm sheath}\sim c\,t_{pp}\sim 1\,\mathrm{mm}$, many orders of magnitude below any reasonable sheath extent. A nonthermal hadron will inevitably interact, and since $\sigma_{pp}$ only grows with energy in the UHECR range, avoiding this constraint requires acceleration at radii where the ambient density has fallen by many additional orders of magnitude.

	\begin{figure}[t!]
		\centering
		\includegraphics[width=\linewidth]{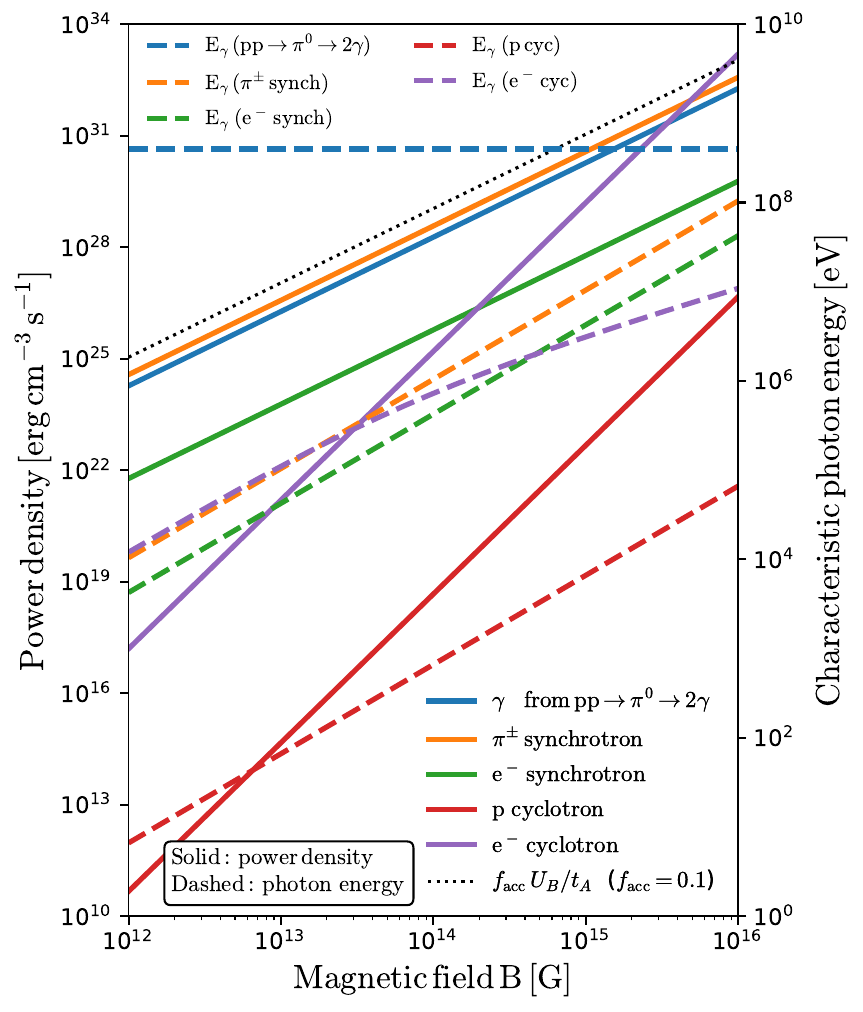}
		\caption{
			Power densities (left axis,
			$\mathrm{erg\,cm^{-3}\,s^{-1}}$) and characteristic photon energies
			(right axis, eV) versus magnetic field, $B$, for the main radiative channels.
			The hadronic channels
			($pp\to\pi^0\to2\gamma$ and $\pi^\pm$ synchrotron) are normalized as
			$P = k\,f_{\rm acc}\,U_B/t_A$ with $f_{\rm acc}=0.1$ and $k=0.17$; the
			dotted black line shows the reference $f_{\rm acc}\,U_B/t_A$. Electron synchrotron uses the supply-limited fraction
			$f_e = m_e/m_p$. Proton and electron cyclotron are computed for
			thermal particles with $\gamma\simeq 1$ and
			$\sin\alpha_p=\sin\alpha_e=10^{-8}$, and are not supply-limited, like their nonthermal counterparts.
			For the fiducial parameters, both pion channels dominate the deposited
			power over the electron synchrotron channel by a factor
			$\sim k_{\pi^0}/f_e \sim 300$. The right axis shows that
			$\pi^0$-decay photons and $\pi^\pm$ synchrotron photons carry
			energies of $\sim 10^8$--$10^9$\,eV and $\sim 10^7$--$10^9$\,eV,
			respectively, over the plotted field range. These energies should be interpreted as characteristic injection energies; the photons need not propagate as physical tower photons because of QED effects.}
		\label{fig:powers}
	\end{figure}

	\paragraph{Radiative-channel prescriptions}
	In Figure~\ref{fig:powers}, we evaluate these channels over a logarithmic magnetic-field grid spanning $10^{12}$--$10^{16}$\,G for a fiducial tower with $\sigma_p=5$, $H=100$\,km, and $v_A=c\sqrt{\sigma_p/(1+\sigma_p)}$.
	At each $B$, we compute $U_B$, the reference dissipation rate $U_B/t_A$, and the density $n_p=B^2/(4\pi \sigma_p m_p c^2)$, with $n_e=n_p$. The $pp\to\pi^0\to2\gamma$ and $\pi^\pm$ synchrotron channels are normalized to the supply-limited hadronic budget in Eq.~\eqref{eq:Phad_supply}. Electron synchrotron is likewise treated as supply-limited with fractional power $f_e\simeq m_e/m_p$, whereas the thermal proton and electron cyclotron channels are computed directly from $n\,P_{\rm cyc,single}$ and are not supply-limited. We also evaluate the characteristic photon energy of each component, using a QED-corrected, burnoff-limited prescription for synchrotron emission (negligible in this energy range), and the lab-frame cyclotron fundamental for the thermal cyclotron channels (Appendix~\ref{app:cyclotron_main},~\ref{app:synch_discussion}, and~\ref{app:synch_all}).

	\paragraph{Pitch-angle sensitivity of cyclotron emission}
	The assumed pitch angle is a key control parameter in this model because the thermal cyclotron emissivity is very sensitive to both magnetic field strength and transverse momentum. At fixed $\sigma_p$, one has $n_e\propto B^2$ and $P_{{\rm cyc},e}\propto B^2\sin^2\alpha_e$, so the corresponding volume emissivity scales as $j_{{\rm cyc},e}\propto B^4\sin^2\alpha_e$, much more steeply than the supply-limited hadronic channels, which scale as $U_B/t_A\propto B^2$. As a result, the condition that electron cyclotron remain subdominant at $B\sim10^{15}$--$10^{16}\,$G is not generic: it requires that $\alpha_e$ stay extremely small over most of the emitting volume. For the adopted values $\sin\alpha_p=\sin\alpha_e=10^{-8}$, both proton and electron cyclotron remain subdominant, but even a modest increase in $\alpha_e$ would strongly amplify the cyclotron contribution without comparably changing its characteristic photon energy, allowing it to overtake the other channels at the high-field end. This also means that any scenario in which cyclotron dominates while still matching the observed short-GRB energy budget requires some degree of fine tuning. Because the cyclotron output rises as $B^4\sin^2\alpha_e$, small changes in either $B$ or $\alpha_e$ can move the total radiated energy by orders of magnitude over a $\sim$second engine duration. The minimum-pitch-angle estimate in Appendix~\ref{app:pitch-ang}, $\sin\alpha_{\rm min}\sim \gamma\rho_0/R_c\lesssim10^{-17}\gamma$ for fiducial inner-tower parameters, helps motivate why such suppression may be achievable, but it also makes clear that cyclotron dominance is not a robust outcome of the model and instead depends on maintaining a narrowly tuned transverse momentum distribution while avoiding overproduction of prompt emission. In that sense, the hierarchy shown in Figure~\ref{fig:powers} should be understood as conditional on efficient radiative damping of perpendicular momentum outside localized heating sites, as expected in strongly magnetized, photon-rich outflows \citep[e.g.,][]{beloborodov+07,gill+14,thompson+14}.
	
	This is a volume-averaged statement, the pitch angle is not necessarily minimal everywhere. Inside and adjacent to reconnecting current sheets, non-ideal electric fields can transiently enhance $\alpha_e$, so leptonic emission is locally stronger in those regions. In the ultra-strong fields considered here, however, electrons with appreciable $\alpha_e$ synchrotron-cool on timescales far shorter than the sheet crossing time, so their large pitch angles are erased before they can sustain a volume-filling, high-energy leptonic cascade. The hadronic channel is not subject to this constraint: $pp$ collisions continuously inject large-angle pions and secondary leptons, replenishing the perpendicular momentum that leptonic processes cannot maintain.

	\paragraph{Energy partition among radiative channels}
	The hadronic power in Eq.~\eqref{eq:Phad_supply} is distributed among channels by their energy fractions: the $\pi^0$-decay $\gamma$-ray
	channel receives a fraction $k_{\pi^0} \simeq 0.17$ per inelastic collision, and charged-pion synchrotron receives a comparable fraction $k_{\pi^\pm}\simeq 0.17$ (by approximate isospin symmetry), so both hadronic channels carry equal power densities scaling as $B^2$. Electron synchrotron is
	also supply-limited and the fraction of reconnection power channelled to electrons is $f_e \simeq m_e/m_p \sim 5\times 10^{-4}$, making it subdominant to the pion channels by a factor $\simeq k_{\pi^0}/f_e \sim 300$. The thermal proton and electron cyclotron components are evaluated separately for $\gamma_p\simeq\gamma_e\simeq1$ and with $\sin\alpha=10^{-8}$.

	\paragraph{Dominant power channels and pair loading}
	Figure~\ref{fig:powers} shows both the power densities and the
	characteristic photon energies of each channel. The supply-limited hadronic and electron-synchrotron components scale as $B^2$ through $U_B/t_A$, whereas the cyclotron channels follow their own direct single-particle emissivities and therefore need not share the same scaling. The pion sector dominates the deposited electromagnetic power over electron synchrotron, confirming that hadronic dissipation is the primary electromagnetic channel for pair loading and high-energy radiation in the inner tower. In particular, pair loading from $pp$ interactions is expected to occur preferentially in regions containing reconnecting current sheets, where nonthermal protons are injected and dissipated.
	
	\paragraph{Connection to short GRB energetics.}
	Sustained over the $\sim 0.1$ to $1\,\mathrm{s}$ duration of a short GRB, the supply-limited hadronic budget of Eq.~\eqref{eq:Phad_supply} yields a total electromagnetic energy of $\gtrsim 10^{50}\,\mathrm{erg}$ for an assumed tower with radius $\sim 10\,\mathrm{km}$ and height $\sim 100\,\mathrm{km}$, comparable to the beaming-corrected prompt release $E_\gamma \simeq 1.6\times10^{50}\,\mathrm{erg}$ and the isotropic-equivalent energy $E_{\gamma,\rm iso}\simeq 2\times10^{51}\,\mathrm{erg}$ inferred for the short-GRB population \citep{fong+15}. This estimate is nevertheless sensitive to the poorly constrained emitting volume: since $E_{\rm had}\propto f_{\rm acc} B^2V$, reducing the tower volume below $\pi(10\,\mathrm{km})^2\times100\,\mathrm{km}$ by a factor $\eta_V$ requires $B\to B/\sqrt{\eta_V}$ to preserve the total energy output. An order-of-magnitude reduction in volume therefore raises the required field from $10^{15}$ to $\sim 3\times10^{15}\,\mathrm{G}$, while a two-order-of-magnitude reduction pushes it to the $\sim 10^{16}\,\mathrm{G}$ upper end suggested by GRMHD simulations \citep{combi+23,Most:2023sft,Kiuchi:2023obe,Bamber:2024kfb,Fields:2025afm}. This scaling also assumes that the magnetic reservoir is dissipated rather than advected out of the tower, making the reconnection efficiency critical. Whether this energy emerges as prompt $\gamma$-rays further depends on where photons are produced relative to the magnetically opaque region. Deep in the tower ($B\gg B_Q$), the $\gamma B\to e^\pm$ threshold $\epsilon\ge 2/\sin\theta_B$ (Eq.~\ref{eq:pair-thr-ground}) defines an escape cone through which only nearly field-aligned photons can free-stream outward \citep{baring+01,hu+19}. Because $\pi^0$-decay photons are emitted isotropically in the pion rest frame and charged pions retain appreciable lab-frame pitch angles for $\gamma_p\lesssim 5$ (Appendix~\ref{app:pion-pitch}), a fraction of the hadronic $\sim 10^8$ to $10^9\,\mathrm{eV}$ photons is injected within this cone and escapes directly, potentially contributing to the sub-GeV to GeV component occasionally observed in short bursts by Fermi-LAT \citep{ackermann+13,ajello+19}. At higher altitudes, where $B\lesssim B_Q$, single-photon conversion shuts off, and pairs injected deeper in the flow radiate classical synchrotron and cyclotron photons that are no longer trapped. The $B\sim B_Q$ surface therefore acts as an effective photospheric boundary for a softer $\sim \mathrm{keV}$ to $\mathrm{MeV}$ leptonic-reprocessed component released at the opacity transition.

	In the inner tower, the dissipated power is deposited primarily into $e^\pm$ pairs through $\gamma B$ conversion of hadronic photons. These pairs subsequently radiate, but the observer-frame spectrum requires radiation transport through anisotropic $\gamma B$ opacity, tower expansion, and synchrotron/Compton cooling at larger radii, which is beyond the scope of this work. We can however identify two expectations: (i) direct photons leaking only from the escape cone $\sin\theta_B<2/\epsilon$ or from altitudes where $B\lesssim B_Q$; and (ii) a reprocessed component from the pair-loaded flow as it expands to lower magnetic opacity, broadly consistent with the short-GRB spectral peak. Their relative weighting depends on the angular structure of the secondaries and the location of the $B\sim B_Q$ surface.

    \paragraph{Connection to neutrino emission.}
Charged-pion decay and the subsequent $\mu\to e\nu_e\nu_\mu$ cascade
populate a nonthermal neutrino band extending up to
$E_\nu^{\rm max}\simeq 60\,\sigma_p\,\mathrm{MeV}$
[Eq.~\eqref{eq:Enu_max}], reaching $\sim 0.3$--$0.6\,\mathrm{GeV}$ for
$\sigma_p=5$--$10$---spectrally distinct from the
$T_\nu\sim 5$--$10\,\mathrm{MeV}$ thermal cooling burst
\citep{perego+17}. A fraction $\eta_\nu\simeq 0.1$ of the
supply-limited hadronic dissipation [Eq.~\eqref{eq:LnuNT_budget}],
integrated over an engine duration $\Delta t\sim 0.1$--$1\,\mathrm{s}$
characteristic of short GRBs, sources this band. Because the
dissipated power tracks the magnetic energy density,
$L_\nu^{\rm NT}\propto U_B\propto B^2$, both the emitted neutrino
energy $E_\nu^{\rm NT}\propto B^2 V\Delta t/t_A$ and the
emitted-neutrino count $N_\nu^{\rm NT}$ scale quadratically with the
tower field. For $N_\nu^{\rm NT}\sim 3\times 10^{52}$ emitted
neutrinos (Eq.~\eqref{eq:NnuNT}), a Hyper-Kamiokande-class detector
\citep{HYPERK18} reaches a horizon
$D_{\rm det}\sim 100$--$200\,\mathrm{kpc}$ at $B=10^{15}\,\mathrm{G}$,
extending to $\sim 1\,\mathrm{Mpc}$ at $B=10^{16}\,\mathrm{G}$: the
detection rate $N_{\rm det}\propto N_\nu^{\rm NT}\,M\,D^{-2}\propto
B^2 V\Delta t\,M\,D^{-2}$ implies a horizon $D_{\rm det}\propto B\,(V\Delta t\,M\,\epsilon)^{1/2}$ that grows linearly with the tower field. A detection would therefore directly measure $\sigma_p$ from the spectral cutoff and constrain the
magnetic-energy reservoir $\eta_\nu f_{\rm acc}B^2V\Delta t/t_A$(Appendix~\ref{app:neutrino_detect}). This estimate should be regarded as close to an upper limit, since large-pitch-angle $\pi^\pm$ can suffer substantial synchrotron losses before decay, reducing the nonthermal neutrino yield (Appendix~\ref{app:neutrino_detect}).

	For $E_\nu \gtrsim 100\,\mathrm{MeV}$, the dominant background is the atmospheric-neutrino flux ($\sim 10^4\,\mathrm{yr}^{-1}$ at Hyper-Kamiokande in the $50$–$600\,\mathrm{MeV}$ band). Discriminating power comes from coincidence with a BNS merger: gravitational-wave and electromagnetic counterparts restrict the search to $\lesssim$ few s and $\lesssim$ deg$^2$, suppressing the effective background by $\sim 10^{10}$ and yielding a near-zero-background counting experiment. The band is also spectrally separated from the thermal remnant burst \citep{Kyutoku:2017wnb,Cusinato:2021zin}---also probed in simulations, e.g., \citealt{Rosswog:2003rv,Sekiguchi:2015dma,Foucart:2015gaa,Radice:2023zlw}--and from TeV–PeV jet/ejecta/fallback/pulsar-wind channels \citep{Kimura:2017kan,Decoene:2019eux,farrar25,Murase:2009pg,Gao:2013rxa,Fang:2017tla,Kimura:2018vvz,Mukhopadhyay:2024ehs}. A coincident $\sim 100$–$600\,\mathrm{MeV}$ excess would signal hadronic tower dissipation, with the cutoff measuring $\sim \sigma_p$ and the fluence constraining $\eta_\nu f_{\rm acc}B^2 V\Delta t/t_A$.
	
	\paragraph{Negligible impact on r-process nucleosynthesis.}
		The hadronic processes analyzed above can strongly modify the local  composition within the magnetic tower itself. 
        Pair loading can yield an $e^\pm$ density far in excess of the baryon density, well above the $Y_e$-regime compatible with r-process nucleosynthesis. 
        The tower is likely not itself a heavy-element  production site, consistent with the established picture in which  r-process nucleosynthesis in BNS mergers takes place in the  equatorial dynamical ejecta and the mid-latitude disk-wind cone,  both angularly separated from the polar funnel.
		
		The emitted nonthermal neutrinos are also not likely to modify  $Y_e$ at these external r-process sites. The nonthermal channel  emits $N_\nu^{\rm NT}\sim 3\times 10^{52}$ neutrinos at  $E_\nu \gtrsim 100\,\mathrm{MeV}$, compared  with $\lesssim 10^{53}\,\mathrm{erg}/(15\,\mathrm{MeV}) \sim 10^{58}$ thermal neutrinos \citep[e.g.,][]{Kyutoku:2017wnb}, and even with the $\sim 100\times$ enhancement of the quasi-elastic cross section $\sigma_{\nu N}\propto E_\nu^2$ \citep{formaggio+12}, the nonthermal  contribution to $Y_e$ modification at the r-process sites is at most  $\sim 10^{-2}$ of the thermal-wind contribution included in 
		standard r-process calculations 
		\citep{wanajo+14,fujibayashi+20,curtis+24}.

	\section{Conclusions}
	\label{sec:conclusions}
	
	We combined GRMHD-motivated BNS post-merger magnetic tower conditions with an analytic treatment of hadronic interactions and strong-field QED processes to assess particle acceleration, pair loading, and high-energy emission. Our main findings are:
	
	(i) \textit{Acceleration.} Magnetic reconnection is the most plausible channel for accelerating protons to mildly relativistic energies in the strongly radiative, small-pitch-angle tower. 
		
		(ii) \textit{Pair loading.} Purely leptonic pair loading---including
		resonant inverse Compton on soft seed photons---is bottlenecked by
		rapid pitch-angle damping and by the tendency of one-photon magnetic
		conversion to populate low Landau levels. Once protons reach
		$\gamma_p\sim\mathrm{few}$, inelastic $pp$ collisions inject
		large-angle pion secondaries that drive
		$\pi^0\to 2\gamma\to e^\pm$ cascades, providing the
		perpendicular-momentum source the leptonic channel lacks. Charged-pion
		synchrotron supplies an additional channel that can dominate over pion
		decay at the highest fields.
		
		(iii) \textit{Electromagnetic output.} The pion sector dominates the injected high-energy power in the inner tower, with the dissipated energy deposited primarily into $e^\pm$ pairs. This sets the energy reservoir available for subsequent radiation but does not by itself fix the observer-frame photon index, which depends on transport processes outside the scope of this work.
		
		(iv) \textit{Multimessenger signature.} The nonthermal hadronic channel produces a neutrino tail at $E_\nu\simeq 60\,\sigma_p\,\mathrm{MeV}\sim0.1$--$0.6\,\mathrm{GeV}$, clearly separated from the thermal cooling emission at $T_\nu\sim5$--$10\,\mathrm{MeV}$. For $B=10^{15}\,\mathrm{G}$, this component would yield a Hyper-Kamiokande few-event horizon of $D_{\rm det}\sim100\,\mathrm{kpc}$ (Appendix~\ref{app:neutrino_detect}). Its cutoff would probe the tower magnetization $\sigma_p$, while its fluence would constrain the magnetic-energy reservoir. This energy range is distinct from both the thermal neutrinos emitted by the hot remnant and the possible TeV-and-above neutrino emission associated with jet--ejecta interactions, fallback accretion, or pulsar-wind interactions in the surrounding nebula.

	Overall, hadronic dissipation appears to be a major driver of both pair loading and high-energy radiation in BNS post-merger magnetic towers, with likely implications for the microphysical state of the outflow and the prompt emission of short GRBs.
	
	Several caveats remain. Our treatment does not self-consistently follow radiation transport and pair cascades in a kinetic framework, and the escape of the resulting $\gamma$-ray emission will depend on the full tower geometry and angular distribution of the secondaries. These questions should be addressed in future work.

	\section{Acknowledgements}
	We would like to thank Eliot Quataert, Anatoly Spitkovsky, Adam Burrows, Lorenzo Sironi, Christopher Thompson, and Ani Prabhu for useful conversations. R.M. is supported by a Lyman Spitzer Fellowship at Princeton University. ERM and JW acknowledge support from NASA's ATP program under grant 80NSSC24K1229 and by the National Science Foundation under grants No. PHY-2309210, and MUSES OAC-2103680. This work made use of Delta at the National Center for Supercomputing Applications (NCSA) through allocation PHY210074 from the Advanced Cyberinfrastructure Coordination Ecosystem: Services \& Support (ACCESS) program, which is supported by National Science Foundation grants \#2138259, \#2138286, \#2138307, \#2137603, and \#2138296. Simulations were also performed on the NSF Frontera supercomputer under grant AST21006. 
	
	\software{ 
		EinsteinToolkit \citep{Loffler:2011ay},
		Frankfurt/IllinoisGRMHD \citep{Most:2019kfe,Etienne:2015cea}
		Lorene \citep{Gourgoulhon:2000nn},
		kuibit \citep{kuibit},
		matplotlib \citep{Hunter:2007},
		numpy \citep{harris2020array},
		scipy \citep{2020SciPy-NMeth}
	}
	
	\balance	
	\appendix
	
	\section{Curvature Drift in BNS Towers}
	\label{app:curv_drift_scales}
	A natural microscopic transverse separation follows from the small
	non-parallel velocity allowed when the total speed remains close to $c$.
	For $v^2=v_\parallel^2+v_\perp^2\simeq c^2$ and
	$v_\parallel=\beta_\parallel c$, one has
	$\beta_\perp\sim(1-\beta_\parallel^2)^{1/2}\sim\gamma^{-1}$ for
	ultra-relativistic, field-aligned particles. Over one curvature time $t_c\sim R_c/v_\parallel$,
	\begin{equation}
		d_\perp\sim v_\perp t_c
		\sim
		\frac{\beta_\perp}{\beta_\parallel}R_c
		\simeq
		10^{-3}
		\left(\frac{R_c}{1\,{\rm km}}\right)
		\left(\frac{10^8}{\gamma\beta_\parallel}\right)
		{\rm cm}.
	\end{equation}
	To achieve order-unity relative displacements across this scale and get energy gain requires perturbations varying across the field on
	$\lambda_\perp\lesssim d_\perp$. For a magnetized jet with $\sigma\simeq5$, the relativistic Alfv\'{e}n speed is
	$v_A=c\sqrt{\sigma/(1+\sigma)}\simeq c$. With Alfv\'{e}nic amplitude $a=\delta B_\perp/B$, the shear-induced relative velocity between lines separated by $d_\perp$ is
	\begin{equation}
		\frac{\Delta v_\perp}{\mathrm{cm\,s^{-1}}}\sim v_A a\frac{d_\perp}{\lambda_\perp}
		\simeq
		2.7\times10^{10}
		a
		\left(\frac{d_\perp}{10^{-3}\mathrm{cm}}\right)
		\left(\frac{10^{-3}\mathrm{cm}}{\lambda_\perp}\right)
	\end{equation}
	Appreciable slippage across $d_\perp\sim10^{-3}\,\mathrm{cm}$ therefore requires both nonnegligible amplitude $a$ and transverse structure on $\lambda_\perp\lesssim10^{-3}\,\mathrm{cm}$, which appears implausibly small.

	\section{Inelastic Proton--Proton Interactions}
	\label{app:pp_inelastic}
	
	The inelastic $pp$ interaction rate is $t_{pp}^{-1}=n_p\sigma_{pp}\kappa_{pp}c$ with $\sigma_{pp}\sim(3$--$4)\times10^{-26}\,\mathrm{cm^2}$ and $\kappa_{pp}\sim0.5$ \citep{murase24}. With $n_p=B^2/(4\pi\sigma_p m_p c^2)$, the neutral-channel energy mapping is $E_\gamma^{(\pi^0)}\simeq b E_p$, $b\approx0.085$, giving $k_{\pi^0\gamma}\approx2b\approx0.17$ for the total $\gamma$-ray fraction \citep{kelner+06,kamae+06}.
	
	The kinematic threshold for single neutral-pion production in the lab frame $pp\to pp\pi^0$ is $T_p^{\rm th}=2m_\pi+m_\pi^2/(2m_p)\simeq0.2797\,\mathrm{GeV}$, i.e.\ $\gamma_p^{\rm th}\simeq1+T_p^{\rm th}/m_p\approx1.3$ \citep{kafexhiu+14}. For $T_p\simeq1$--$3\,\mathrm{GeV}$ ($\gamma_p\simeq2.1$--$4.2$), the inelastic cross section is $\sigma_{\rm inel}\sim20$--$30\,\mathrm{mb}\simeq(2$--$3)\times10^{-26}\,\mathrm{cm^2}$ with only a weak energy dependence \citep{kafexhiu+14}.
	
	The average neutral-pion multiplicity for $1\,\mathrm{GeV}\le T_p<5\,\mathrm{GeV}$ is fitted by \citet{kafexhiu+14} as $\langle n_{\pi^0}\rangle=-6\times10^{-3}+0.237Q_p-0.023Q_p^2$ (their Eq.~6), where $Q_p=(T_p-T_p^{\rm th})/m_p$. This yields $\langle n_{\pi^0}\rangle\simeq0.16$, $0.35$, $0.49$ at $T_p=1$, $2$, $3\,\mathrm{GeV}$. The total charged-pion multiplicity is approximately $2\langle n_{\pi^0}\rangle$ at these energies, based on approximate isospin symmetry in the multi-pion regime.
	
	The angular distribution is similarly well controlled in this mildly relativistic regime. \citet{kafexhiu+14} adopt isotropic $\pi^0$ emission in the $pp$ center-of-momentum frame for $T_p<1,\mathrm{GeV}$ and show that this approximation reproduces the data to better than $\sim10\%$. The complementary resonance treatment of \citet{kamae+06}, which also assumes isotropic CM pion emission, reproduces the measured $\pi^0$ kinetic-energy spectra at $T_p=0.65$--$2.0\,\mathrm{GeV}$ to within $\sim20\%$. Thus, for the mildly relativistic protons relevant here, $\gamma_p\lesssim4$--5, the pion distribution can be taken to be approximately isotropic in the CM frame.
	
	\section{Coulomb Collisions and Collisionality of the Tower}
	\label{app:coulomb}
	
	\subsection{Proton thermalization time}
	The Coulomb collision time is:
	\begin{equation}
		t_C \;=\; \frac{3}{4\sqrt{\pi}}\,
		\frac{(k_BT)^{3/2}\,m_p^{1/2}}{n_p\,e^4\,\ln\Lambda},
		\label{eq:tC_def}
	\end{equation}
	or,
	\begin{equation}
		\frac{t_C}{s} \;\simeq\; 2\!\times\!10^{-16}\!\left(\frac{\sigma_p}{5}\right)\!
		\left(\frac{B}{10^{15}\,\mathrm{G}}\right)^{\!-2}\!
		\left(\frac{T}{1\,\mathrm{MeV}}\right)^{\!3/2}\!
		\left(\frac{\ln\Lambda}{10}\right)^{\!-1}
		\label{eq:tC_main}
	\end{equation}
	The ratio to inelastic $pp$ (Appendix~\ref{app:pp_inelastic}) is:
	\begin{equation}
		\frac{t_C}{t_{pp}} \;\simeq\; 0.94\!
		\left(\frac{T}{1\,\mathrm{MeV}}\right)^{\!3/2}\!
		\left(\frac{\sigma_{pp}\kappa_{pp}}{1.5\!\times\!10^{-26}\,\mathrm{cm^2}}\right)\!
		\left(\frac{\ln\Lambda}{10}\right)^{\!-1}
		\label{eq:tC_tpp_ratio}
	\end{equation}
	Numerically: $t_C/t_{pp}\simeq 1$ at $T=1\,\mathrm{MeV}$, $\simeq 10$ at $T=5\,\mathrm{MeV}$, and $\simeq 30$ at $T=10\,\mathrm{MeV}$. In the cool bulk, Coulomb relaxation and inelastic $pp$ run at comparable rates, maintaining a Maxwellian background at temperatures below the pion production threshold $T_p^{\rm th}\simeq 0.28\,\mathrm{GeV}$. In hot reconnection sheets, $pp$ dominates by an order of magnitude or more, so the nonthermal proton tail produced by reconnection deposits its energy into pions before being Coulomb-isotropized.
	
	\subsection{Reconnection geometry and current-sheet thickness}
	\label{subsec:collisionless}
	
	The relevant criterion for reconnection physics is not whether the bulk plasma is collisional in the MHD sense, but whether the Coulomb mean free path exceeds the sheet thickness, set by the electron skin depth
	\citep{uzdensky+10}. With
	$d_e=c/\omega_{pe}=(m_e c^2/4\pi n_e e^2)^{1/2}\simeq
	1.6\times 10^{-10}\,\mathrm{cm}$ at $n_e=n_p\simeq 10^{31}\,\mathrm{cm^{-3}}$,
	the corresponding mean free paths are
	$\lambda_C=v_{\rm th}\,t_C\simeq 5\times 10^{-6}$~cm at $T=5\,\mathrm{MeV}$, giving
	\begin{equation}
		\frac{\lambda_C}{d_e}\;\simeq\;3\!\times\!10^{4}\!\left(\frac{T}{5\,\mathrm{MeV}}\right)^{\!2}\!\left(\frac{\sigma_p}{5}\right)^{\!-1/2}\!\left(\frac{B}{10^{15}\,\mathrm{G}}\right)^{\!-1}\!.
		\label{eq:lambda_de}
	\end{equation}
	(The $T^2$ scaling comes from $\lambda_C=v_{\rm th}\,t_C\propto T^{1/2}\,T^{3/2}$;
	the density dependence of $d_e$ adds the $\sigma_p^{-1/2}B^{-1}$ piece.)

	The current sheets identified in Fig.~\ref{fig:fig2_reconnection_site} are macroscopic reconnection sites whose resolved thickness is set by the GRMHD grid, $\Delta x\simeq214\,{\rm m}$, not by the physical dissipation scale. For a representative sheet length $L\sim10\,{\rm km}$, the Sweet--Parker thickness at the plasmoid threshold is $\delta_{\rm pl}\sim L/S_c^{1/2}\sim100\,{\rm m}$, where $S_c\sim10^4$ is the critical Lundquist number above which elongated resistive-MHD sheets become tearing unstable \citep{loureiro+07,bhattacharjee+09,uzdensky+10}. Thus the resolved sheets are already at the scale where a physical sheet would be expected to fragment. The relevant kinetic scales are much smaller:
		$d_e\simeq1.6\times10^{-10}\,{\rm cm}$ and
		$d_i\simeq7\times10^{-9}\,{\rm cm}$ for
		$n_e\simeq10^{31}\,{\rm cm^{-3}}$.
		
	Whether this fragmentation terminates collisionally or collisionlessly is set by the microscopic magnetic diffusivity $\eta_m$. Since the GRMHD sheet has temperature $T\sim5\,{\rm MeV}$, the current-carrying electrons are relativistically hot, with $k_B T/m_e c^2\sim10$. We therefore estimate the magnetic diffusivity as $\eta_m=c^2/(4\pi\sigma_{\rm cond})$, using the relativistic collisional conductivity $\sigma_{\rm cond}\sim n_e e^2/(\bar{\gamma}_e m_e\nu_C)$:
		\begin{equation}
			\eta_m\sim d_e^2\nu_C
			\sim c d_e\left(\frac{d_e}{\lambda_C}\right),
		\end{equation}
		where $\nu_C\sim c/\lambda_C$ and $\bar{\gamma}_e \sim \text{a few}$. One finds $\eta_m\sim 10^{-4}\,{\rm cm^2\,s^{-1}}$ based on Equation~\eqref{eq:lambda_de} and hence $S=L v_A/\eta_m\sim2\times10^{20}\gg S_c$ for $L\sim10\,{\rm km}$ and $v_A\sim c$. A purely resistive cascade reaches local marginality at
		\begin{equation}
			\delta_c^{\rm rel}\sim S_c^{1/2}\frac{\eta_m}{c}
			\sim5\times10^{-13}\,{\rm cm},
		\end{equation}
		which lies below $d_e$, while we also have $\lambda_C\gg d_e$. The scale ordering is therefore $\delta_{\rm pl}\gg d_i\gg d_e\gg \delta_c^{\rm rel}$. The physical dissipation layer is consequently collisionless and regulated by ion and electron kinetic physics. This is the relevant condition for proton acceleration and interactions. Protons entering these fragmented sheets experience collisionless reconnection, with gyroradii at $\gamma_p\sim{\rm few}$ that are comparable to the kinetic scales for $B\sim10^{15}$--$10^{16}\,{\rm G}$.

	\section{Cyclotron Emission in the Strong-Field Limit}
	\label{app:cyclotron_main}
	
	For perpendicular velocity $\beta_\perp c$ and species $s$:
	\begin{align}
		P_{\rm cyc,s}&=\frac{2q_s^4B^2}{3m_s^2c^3}\beta_\perp^2=\frac{4}{3}\sigma_{T,s}cU_B\beta_\perp^2,\\
		\epsilon_{\rm cyc}^{\rm rest}&=\hbar\frac{q_sB}{m_sc}.
	\end{align}
	For $B\gtrsim B_Q$ the Landau spacing gives
	\begin{equation}
		\epsilon_{e,\rm cyc}^{\rm rest}=m_ec^2\left(\sqrt{1+2B/B_Q}-1\right),
	\end{equation}
	producing MeV photons at $B\sim10^{15}\,\mathrm{G}$.
	
	For a photon of dimensionless energy $\epsilon=E_\gamma/(m_ec^2)$ propagating at angle $\theta$ to $\mathbf{B}$, one-photon magnetic pair creation is constrained by energy--momentum conservation into discrete Landau states \citep{harding+06}:
	\begin{align}
		E_n+E_{n'}&=\epsilon,\quad p+q=\epsilon\cos\theta,\\
		E_n&=\sqrt{1+p^2+2nB'},\quad E_{n'}=\sqrt{1+q^2+2n'B'},
	\end{align}
	where $p$ and $q$ are the parallel momenta and $B'=B/B_Q$. The lowest free-pair threshold ($k$-polarization, ground states $(0,0)$) is \citep{harding+06}
	\begin{equation}
		\epsilon\ge\frac{2}{\sin\theta}.
		\label{eq:pair-thr-ground}
	\end{equation}
	
	\paragraph{Cyclotron photon energy and the pitch-angle requirement.}
	The fundamental cyclotron spacing is
	\begin{equation}
		\epsilon_{\rm cyc}\equiv\frac{E_1-E_0}{m_ec^2}=\sqrt{1+2B'}-1.
		\label{eq:epscyc}
	\end{equation}
	For $B=10^{15}\,\mathrm{G}$ ($B'\simeq22.7$): $\epsilon_{\rm cyc}\simeq5.81$, $E_\gamma\simeq2.97\,\mathrm{MeV}$. For $B=10^{16}\,\mathrm{G}$ ($B'\simeq227$): $\epsilon_{\rm cyc}\simeq20.3$, $E_\gamma\simeq10.4\,\mathrm{MeV}$. The minimum angle for $\gamma B$ conversion of a cyclotron photon is $\sin\theta_{\min}=2/\epsilon_{\rm cyc}$.
	
	\paragraph{When are produced pairs in $n>0$?}
	For at least one lepton to be created with $n>0$ (required for subsequent Landau de-excitation), the channel $(0,1)$ threshold must be met in the perpendicular frame:
	\begin{align}
		\epsilon_\perp&\equiv\epsilon\sin\theta\ge\epsilon_{\perp,\rm thr}^{(0,1)}\\
		\epsilon_{\perp,\rm thr}^{(0,1)}&=1+\sqrt{1+2B'}
		\label{eq:epsperp_thr01}
	\end{align}
	For a fundamental cyclotron photon, $\epsilon\le\epsilon_{\rm cyc}=\sqrt{1+2B'}-1$, so even at $\sin\theta=1$:
	\begin{equation}
		\epsilon_\perp^{\rm max}=\epsilon_{\rm cyc}=\sqrt{1+2B'}-1<1+\sqrt{1+2B'},
	\end{equation}
	i.e.\ the fundamental cyclotron photon falls short of the $(0,1)$ threshold by exactly 2 (in units of $m_ec^2$). Therefore a fundamental cyclotron photon converting via $\gamma B$ can only do so through the ground-state $(0,0)$ channel \citep{harding+06}. Ground-state pairs cannot produce a subsequent MeV de-excitation branch, confirming that the cascade multiplicity for cyclotron-only injection is $\mathcal{M}_{\rm cas}\le1$. This is consistent with detailed cascade simulations at high field when inverse Compton scattering is inactive \citep{medin+10,daugherty+83}.
	
	\paragraph{Cascade weakness at high $B$ and the role of pitch angle.}
	In \citet{medin+10}, photon-initiated cascades are intrinsically weak in the high-field regime because $\gamma B\to e^\pm$ typically populates $(0,0)$ or $(0,1)$ states, leaving little transverse excitation. This is reinforced by the geometric assumption of near-tangent emission, which forces conversion near threshold. Only when a nonnegligible pitch angle $\alpha$ is maintained at emission---so that $\epsilon_\perp=\epsilon\sin\theta$ is large---can higher Landau channels be accessed and the cascade multiplicity be increased. Sustaining such pitch angles generally requires continuous perpendicular momentum injection, as can be provided by $pp$ interactions.
	
	\section{Minimum Pitch Angle}
	\label{app:pitch-ang}
	
	In a curved field with radius $R_c$, the adiabatic-invariant correction gives a minimum perpendicular momentum $p_\perp\sim p_\parallel^2c/(q_sB_0R_c)$, so for ultra-relativistic particles:
	\begin{equation}
		\sin\alpha_{\rm min}\sim\frac{p_\perp}{p_\parallel}\sim\frac{\gamma\rho_0}{R_c},\qquad\rho_0\equiv\frac{m_sc^2}{q_sB_0}.
		\label{eq:alpha_min}
	\end{equation}
	For electrons at $B_0=10^{15}\,\mathrm{G}$ and $R_c\sim10^5\,\mathrm{cm}$:
	\begin{equation}
		{\alpha_{\rm min} \approx}\sin\alpha_{\rm min}\sim\gamma\frac{1.7\times10^{-12}}{10^5}\lesssim10^{-17}\gamma.
	\end{equation}
	
	\section{Pion Pitch-Angle Suppression at High Proton Lorentz Factor}
	\label{app:pion-pitch}
	
	For a projectile proton with lab-frame $\gamma_p$ colliding with a stationary target proton, the CM frame moves with
	\begin{equation}
		\beta_{\rm cm}=\frac{\sqrt{\gamma_p^2-1}}{\gamma_p+1},\qquad\Gamma_{\rm cm}=(1-\beta_{\rm cm}^2)^{-1/2}.
	\end{equation}
	If a pion is emitted in the CM frame with speed $\beta'_\pi c$ at CM angle $\theta'$, its lab-frame pitch angle satisfies
	\begin{equation}
		\sin\alpha_\pi=\frac{\beta'_\pi\sin\theta'}{\sqrt{(\beta'_\pi\sin\theta')^2+\Gamma_{\rm cm}^2(\beta'_\pi\cos\theta'+\beta_{\rm cm})^2}}.
	\end{equation}
	In the relativistic limit $\beta'_\pi\simeq1$ this reduces to the standard aberration formula and shows that $\sin\alpha_\pi$ decreases with increasing $\gamma_p$ for fixed $\theta'$. Since the efficiency of $\gamma B\to e^\pm$ depends on $\chi_\gamma\propto(E_\gamma/m_ec^2)(B/B_Q)\sin\theta_B$, this kinematic beaming reduces the effective pair-loading efficiency of $\pi^0$-decay photons at higher proton energies.
	
	\section{Synchrotron Energetics}
	\label{app:synch_discussion}
	
	\subsection{Classical synchrotron and burnoff}
	
	With $B_\perp\equiv B\sin\alpha$, acceleration power $P_{\rm acc}=e\xi Bc$ (where $\xi\equiv \mathcal{E}/B$, where $\mathcal{E}$ is the electric field), and synchrotron loss power $P_{\rm rad}=g(\chi)(4/3)\sigma_TcU_B\gamma^2\sin^2\alpha$, the burnoff condition $P_{\rm acc}=P_{\rm rad}$ gives:
	\begin{equation}
		\gamma_{\rm bo}^2g(\chi)=\frac{6\pi e\xi}{\sigma_TB\sin^2\alpha}.
	\end{equation}
	In the classical limit $g\to1$, the burnoff-limited peak photon energy is
	\begin{equation}
		E_{\gamma\rm burn}=\frac{0.98}{\alpha_f}\frac{\xi}{\sin\alpha}m_ec^2,
	\end{equation}
	independent of $B$.
	
	\subsection{QED synchrotron reduction and deep-quantum burnoff}
	
	The quantum correction to synchrotron losses is \citep{blackburn+14}
	\begin{equation}
		g(\chi)=\left(1+4.8\chi+1.7\chi^2\right)^{-2/3}.
	\end{equation}
	In the deep-quantum limit $\chi\gg1$, $g(\chi)\simeq C_q\chi^{-4/3}$ with $C_q=1.7^{-2/3}$, giving
	\begin{equation}
		\gamma_{\rm bo}(B)=\frac{1}{\sin\alpha\,B_Q^2}\left(\frac{6\pi e\xi}{\sigma_TC_q}\right)^{3/2}B^{1/2}\propto B^{1/2}.
	\end{equation}
	The effective characteristic photon energy at quantum burnoff is $E_\gamma\simeq\min[E_{\gamma q}(\gamma_{\rm bo}),E_{\gamma\rm max}^{(\rm bo)}]$ where $E_{\gamma q}=0.44(\gamma^2B_\perp/B_Q)m_ec^2$ and $E_{\gamma\rm max}^{(\rm bo)}=\gamma_{\rm bo}m_ec^2$ \citep{bell08,blackburn+14}.
	
	The QED power reduction factor used in power density estimates is
	\begin{equation}
		G(\chi)=\left[1+4.8(1+\chi)\ln(1+1.7\chi)+2.44\chi^2\right]^{-2/3},
	\end{equation}
	so $P_{\syn,s}^{\rm QED}=P_{\syn,s}^{\rm class}G(\chi)$.

	\section{Non-thermal Neutrino Emission and Detection Prospects}
	\label{app:neutrino_detect}
	
	\subsection{Spectral signature}
	
	For the two-body decay $\pi^\pm\to\mu^\pm\nu_\mu$, the neutrino energy in the pion rest frame is
	\begin{equation}
		E_\nu^\ast=
		\frac{m_\pi^2-m_\mu^2}{2m_\pi}
		\simeq 29.8\,{\rm MeV}.
	\end{equation}
	If charged pions are produced with $\gamma_\pi\simeq\sigma_p$ and decay before appreciable synchrotron cooling, the lab-frame cutoff is
	\begin{equation}
		E_\nu^{\rm max}\simeq
		2\gamma_\pi E_\nu^\ast
		\simeq
		60\,\sigma_p\,{\rm MeV}.
		\label{eq:Enu_max}
	\end{equation}
	The subsequent decay $\mu^\pm\to e^\pm\nu_e\nu_\mu$ adds a broader component with characteristic energies
	$E_\nu\sim35\,\sigma_p\,{\rm MeV}$. Thus the hadronic channel populates
	\begin{equation}
		E_\nu\sim30\,{\rm MeV}
		\text{--}
		60\,\sigma_p\,{\rm MeV},
	\end{equation}
	reaching $\sim0.3$--$0.6\,{\rm GeV}$ for $\sigma_p=5$--$10$.
	
	This cutoff is cleanest in the decay-dominated regime. More generally, define
	$B_{\rm syn,\pi^\pm}$ by
	\begin{equation}
		t_{\rm syn,\pi^\pm}(B_{\rm syn,\pi^\pm},\gamma_\pi,\alpha_\pi)
		=
		\gamma_\pi\tau_\pi ,
	\end{equation}
	where $\tau_\pi$ is the charged-pion lifetime and $\alpha_\pi$ is the pion pitch angle. For
	$B\gtrsim B_{\rm syn,\pi^\pm}$, synchrotron radiation damps the perpendicular momentum before decay. Since the loss force acts primarily on $p_\perp$, the pitch angle decreases during cooling while the parallel momentum is comparatively preserved. The decay neutrino energy is then controlled by the residual parallel Lorentz factor, $\gamma_{\parallel,\pi}$, rather than by the initial $\gamma_\pi\simeq\sigma_p$, and Equation~\eqref{eq:Enu_max} becomes an upper envelope. Therefore the cutoff directly measures $\sigma_p$ only when
	\begin{equation}
		t_{\rm dec,\pi^\pm}<t_{\rm syn,\pi^\pm},
	\end{equation}
	which can still hold in $B\sim10^{15}$--$10^{16}\,{\rm G}$ regions if the produced pions have sufficiently small pitch angles.
	
	\subsection{Non-thermal energy budget}
	
	The nonthermal neutrino luminosity is taken to be a fraction $\eta_\nu \simeq 0.1$ of the supply-limited hadronic power,
	\begin{equation}
		L_\nu^{\rm NT}
		\simeq
		\eta_\nu f_{\rm acc}\frac{U_B V}{t_A}
		\label{eq:LnuNT_budget}
	\end{equation}
	where $f_{\rm acc} =0.1$, $V$ is the tower volume, and $t_A$ is the Alfvén crossing time. Numerically,
	\begin{equation}
		L_\nu^{\rm NT}
		\simeq
		10^{49}
		\left(\frac{B}{10^{15}\,{\rm G}}\right)^2
		\left(\frac{V}{V_0}\right)
		\left(\frac{t_A}{1\,{\rm ms}}\right)^{-1}
		{\rm erg\,s^{-1}}
		\label{eq:LnuNT_num}
	\end{equation}
	with $V_0=\pi(10\,{\rm km})^2(100\,{\rm km})$.
	The emitted nonthermal neutrino energy over an engine duration $\Delta t$ is therefore
	\begin{equation}
		E_\nu^{\rm NT}
		\simeq
		10^{49}
		\left(\frac{B}{10^{15}\,{\rm G}}\right)^2
		\left(\frac{V}{V_0}\right)
		\left(\frac{\Delta t}{1\,{\rm s}}\right)
		\left(\frac{t_A}{1\,{\rm ms}}\right)^{-1}
		{\rm erg}
		\label{eq:EnuNT_B}
	\end{equation}
	Equivalently, for a characteristic nonthermal neutrino energy $\langle E_\nu\rangle$,
	\begin{equation}
		N_\nu^{\rm NT}
		\simeq
		3\times10^{52}
		\left(\frac{E_\nu^{\rm NT}}{10^{49}\,{\rm erg}}\right)
		\left(\frac{\langle E_\nu\rangle}{300\,{\rm MeV}}\right)^{-1}
		\label{eq:NnuNT}
	\end{equation}

	\subsection{Detection prospects}
	
	The expected charged-current event yield is
	\begin{equation}
		N_{\rm det}
		\simeq
		\frac{N_\nu^{\rm NT}}{4\pi D^2}
		\sigma_{\nu N} \frac{M}{m_p} \epsilon ,
	\end{equation}
	where $D$ is the source distance, $M$ is the detector of fiducial mass, $m_p$ the mass of a proton target, and $\epsilon_{\rm det}$ is the analysis efficiency. With $\sigma_{\nu N} = 2\times10^{-39}\,{\rm cm^2}$ \citep{formaggio+12},
	\begin{equation}
		N_{\rm det}\simeq
		2
		\left(\frac{N_\nu^{\rm NT}}{10^{52}}\right)
		\left(\frac{M}{187\,{\rm kt}}\right)
		\left(\frac{D}{100\,{\rm kpc}}\right)^{-2}
		\epsilon_{\rm det}
		\label{eq:Ndet}
	\end{equation}
	Substituting the magnetic-energy normalization gives
	\begin{align}
		N_{\rm det}\simeq
		5
		\left(\frac{B}{10^{15}\,{\rm G}}\right)^2
		\left(\frac{V}{V_0}\right)
		\left(\frac{\Delta t}{1\,{\rm s}}\right)
		\left(\frac{t_A}{1\,{\rm ms}}\right)^{-1}\\
		\times \left(\frac{\langle E_\nu\rangle}{300\,{\rm MeV}}\right)^{-1}
		\left(\frac{M}{187\,{\rm kt}}\right)
		\left(\frac{D}{100\,{\rm kpc}}\right)^{-2}
		\epsilon_{\rm det}
		\label{eq:Ndet_B}
	\end{align}
	This expression results in $N_{\rm det}\propto B^2\,V\,\Delta t\,M\,D^{-2}$ and $D_{\rm det}\propto B\, \left(V\Delta t M\epsilon_{\rm det} \right)^{1/2}$.
	
	Among sub-GeV-sensitive facilities, Hyper-Kamiokande \citep{HYPERK18} provides the largest near-future target mass, DUNE offers superior $\nu_e$ imaging with a smaller liquid-argon mass \citep{DUNE20}, JUNO provides high light yield in scintillator \citep{JUNO16}, and Super-Kamiokande remains the current water-Cherenkov benchmark \citep{SUPERK03}. 
	
	\subsection{What a detection constrains}
	
	A neutrino detection above $E_\star\simeq70$--$100\,{\rm MeV}$ would isolate the hadronic component from the thermal cooling burst. The cutoff $E_\nu^{\rm max}\simeq60\sigma_p\,{\rm MeV}$ constrains the proton magnetization at the dissipation site. The fluence normalization, together with the source distance, constrains $\eta_\nu f_{\rm acc}B^2V\Delta t/t_A$, so it measures the magnetic-energy reservoir. Inferring $B$ then requires independent constraints on the emitting volume, duration, acceleration efficiency, baryon loading, and pitch-angle-dependent cooling. Thus, the spectrum probes $\sigma_p$, while the normalization probes the magnetic-energy budget.

	\section{General Synchrotron Volume Emissivities}
	\label{app:synch_all}
	
	For species $s$ with magnetization $\sigma_s=B^2/(4\pi n_s m_sc^2)$, $\sigma_{T,s}=\frac{8\pi}{3}(q_s^2/m_sc^2)^2$ and density $n_s=B^2/(4\pi\sigma_s m_sc^2)$:
	\begin{align}
		P_{\syn,s}(\gamma_s)&=\frac{4}{3}\sigma_{T,s}cU_B\gamma_s^2,\\
		J_s&=n_sP_{\syn,s}=\frac{\sigma_{T,s}}{24\pi^2}\frac{B^4}{\sigma_s m_sc}\gamma_s^2.
	\end{align}
	Characteristic photon energy: $\epsilon_s\simeq\frac{3h|q_s|}{4\pi m_sc}B\gamma_s^2$.
	
	\paragraph{Pion synchrotron.}
	Replacing $m_e\to m_\pi$ and $\sigma_T\to\sigma_T(m_\pi)$: 
	\begin{align}
		t_{\syn,\pi^\pm}&=\frac{6\pi m_\pi^3 {c}}{\sigma_Tm_e^2B^2\gamma_\pi},\qquad
		\epsilon_{\syn,\pi^\pm}\simeq0.44\hbar\frac{eB}{m_\pi c}\gamma_\pi^2.
	\end{align}
	Since $B_{Q,\pi}=B_Q(m_\pi/m_e)^2\simeq3\times10^{18}\,\mathrm{G}$ far exceeds the fields considered here, pion synchrotron remains classical throughout.
	\label{app:pion_synch}

	\section{Pion Decay vs.\ Synchrotron Cooling}
	\label{app:pion_decay_compete}
	
	Charged and neutral pion rest lifetimes are $\tau_{\pi^\pm}\simeq2.6033\times10^{-8}\,\mathrm{s}$ and $\tau_{\pi^0}\simeq8.4\times10^{-17}\,\mathrm{s}$. In the lab frame $t_{\rm dec,\pi^{\pm}}=\gamma_\pi\tau_{\pi^\pm}$. For $B\sim10^{15}\,\mathrm{G}$ and $\gamma_\pi\sim\mathrm{few}$--$10$, $t_{\syn,\pi^{\pm}}\ll t_{\rm dec,\pi^{\pm}}$, so $\pi^0$ decay proceeds promptly to $2\gamma$ while $\pi^\pm$ cool primarily by synchrotron. The condition $t_{\syn,\pi^{\pm}}=t_{\rm dec,\pi^{\pm}}$ for $\sin\alpha_\pi\simeq0.5$ sets the crossover field above which pion synchrotron dominates; from the expressions in Appendix~\ref{app:synch_all} this occurs at $B\sim\mathrm{few}\times10^{15}\,\mathrm{G}$ for $\gamma_\pi\sim2$--5, consistent with Figure~\ref{fig:timescales}.

	\section{Single-Photon Pair Creation ($\gamma B\to e^\pm$) and Attenuation Time}
	\label{app:gamma_to_pairs_B}
	
	Define $\chi=E_\gamma/(2m_ec^2)\times B/B_{\rm cr,e}$ with $B_{\rm cr,e}\simeq4.41\times10^{13}\,\mathrm{G}$. For $\chi\gg1$, the \citet{erber66} approximation gives the effective cross section:
	\begin{align}
		\sigma(\chi)\simeq\frac{1}{2}\frac{\alpha_f}{\lambda_c}\frac{B}{B_{\rm cr,e}}T(\chi),\qquad T(\chi)\simeq0.6\chi^{-1/3},
	\end{align}
	with $\lambda_c=\hbar/(m_ec)$. The mean free path and attenuation time are
	\begin{equation}
		\lambda\simeq\frac{2\lambda_c}{\alpha_f}\frac{B_{\rm cr,e}}{B}\frac{1}{T(\chi)},\qquad t_{\gamma\to e^\pm}=\lambda/c,
	\end{equation}
	consistent with \citet{erber66,daugherty+83}. When additional channels are relevant, $t_{\gamma,\rm dec}^{-1}=t_{\gamma\to e^\pm}^{-1}+t_{\gamma\gamma\to e^\pm}^{-1}+t_{\rm esc}^{-1}$.

	
		\section{Magnetic inverse-Compton scattering}
		\label{app:rics}
		
		In the inner magnetic tower, IC scattering is described by the magnetic Compton cross section $\sigma_{\rm mag}$ rather than by the Thomson cross section. The cross section depends on $B$, the photon energy and direction in the electron rest frame, polarization, Landau-state structure, and the resonant width of the intermediate electron state \citep{daugherty+86,gonthier+14,wadiasingh+18}. We therefore use $\langle\sigma_{\rm mag}\rangle$ only as a local spectrum- and angle-averaged quantity.
		
		For an electron moving along the magnetic field, the seed photon energy
		in the electron rest frame is
		\begin{equation}
			\epsilon_s'=\gamma_e\epsilon_s(1-\beta_e\mu_s),
		\end{equation}
		where $\mu_s$ is the cosine of the angle between the seed photon and the
		electron velocity. Resonant magnetic Compton scattering occurs when $\epsilon_s'$ matches the fundamental Landau spacing
		(cf.~Appendix~\ref{app:cyclotron_main}),
		\begin{equation}
			\epsilon_s'=\sqrt{1+2B'}-1 ,
			\label{eq:rics_res_condition}
		\end{equation}
		with $B'\equiv B/B_Q$. For head-on photons this gives
		\begin{equation}
			\gamma_{\rm res}
			\simeq
			\frac{\epsilon_s'}{2\epsilon_s}
			\simeq
			15
			\left(\frac{\epsilon_s'}{5.81}\right)
			\left(\frac{E_s}{100\,{\rm keV}}\right)^{-1} ,
			\label{eq:gamma_res}
		\end{equation}
		where the normalization corresponds to $B=10^{15}\,\mathrm{G}$. Thus $\sim$MeV seed photons can resonate with mildly relativistic pairs ($\gamma_{\rm res}\sim$ few) at this field, whereas softer X-ray photons require larger $\gamma_e$ or more favorable collision angles.
		
		For pair loading, the relevant question is not only whether a photon is upscattered but whether the upscattered photon's transverse energy is high enough to populate excited Landau states, since ground-state pair
		production does not drive a subsequent cascade
		(Appendix~\ref{app:cyclotron_main}; \S~\ref{sec:qed_pairs}). The
		$(0,0)$ threshold is
		\begin{equation}
			\epsilon_{\rm IC}\sin\theta_{\gamma B}\gtrsim2 ,
			\label{eq:ic_pair_condition_app}
		\end{equation}
		while the $(0,1)$ threshold is
		$\epsilon_{\rm IC}\sin\theta_{\gamma B}\gtrsim 1+\sqrt{1+2B'}$
		(Eq.~\ref{eq:epsperp_thr01}). 
		
		In resonant magnetic Compton scattering, the emitted photon energy in the electron rest frame is of order the Landau spacing, $\epsilon'\simeq\sqrt{1+2B'}-1$. Since the rest-frame emission is concentrated near $\theta'\sim\pi/2$, transformation to the lab frame gives $\epsilon_{\rm IC}\simeq\gamma_e\epsilon'$ and $\sin\theta_{\gamma B}\simeq\gamma_e^{-1}$. The transverse photon energy is therefore approximately Lorentz invariant, $\epsilon_{\rm IC}\sin\theta_{\gamma B}\simeq\epsilon'\sin\theta'\simeq\sqrt{1+2B'}-1$, independent of $\gamma_e$.
		
		This exceeds the $(0,0)$ threshold by a comfortable margin at $B\gtrsim B_Q$, but falls exactly $2$ units below the $(0,1)$ threshold. Resonant IC therefore injects pairs into the ground state and the cascade multiplicity is bounded at $\mathcal{M}_{\rm cas}\le 1$, the same Landau-level bottleneck that suppresses cyclotron-only pair injection (Appendix~\ref{app:cyclotron_main}).

	\balance
	\bibliography{Total,elias_bib,Total_cont}
	\bibliographystyle{aasjournalv7.bst}

\end{document}